# Self-consistent calculation of spin transport and magnetization dynamics

Kyung-Jin Lee[a,b,c,d*], M. D. Stiles[c], Hyun-Woo Lee[e], Jung-Hwan Moon[a], Kyoung-Whan Kim[e], Seo-Won Lee[a]

[a]Department of Materials Science and Engineering, Korea University, Seoul 136-713, Korea

[b]KU-KIST School of Converging Science and Technology, Korea University, Seoul 136-713, Korea

[c]Center for Nanoscale Science and Technology, National Institute of Standards and Technology, Gaitherburg, Maryland 20899, USA

[d]Maryland Nanocenter, University of Maryland, College Park, Maryland 20742, USA

[e]Department of Physics, Pohang University of Science and Technology, Pohang, Kyungbuk 790-784, Korea

* Corresponding author

   E-mail address: kj_lee@korea.ac.kr (K. –J. Lee)

**Abstract**

A spin-polarized current transfers its spin-angular momentum to a local magnetization, exciting various types of current-induced magnetization dynamics. So far, most studies in this field have focused on the direct effect of spin transport on magnetization dynamics, but ignored the feedback from the magnetization dynamics to the spin transport and back to the magnetization dynamics. Although the feedback is usually weak, there are situations when it can play an important role in the dynamics. In such situations, simultaneous, self-consistent calculations of the magnetization dynamics and the spin transport can accurately describe the feedback. This review describes in detail the feedback mechanisms, and presents recent progress in self-consistent calculations of the coupled dynamics. We pay special attention to three representative examples, where the feedback generates non-local effective interactions for the magnetization after the spin accumulation has been integrated out. Possibly the most dramatic feedback example is the dynamic instability in magnetic nanopillars with a single magnetic layer. This instability does not occur without non-local feedback. We demonstrate that full self-consistent calculations generate simulation results in much better agreement with experiments than previous calculations that addressed the feedback effect approximately. The next example is for more typical spin valve nanopillars. Although the effect of feedback is less dramatic because even without feedback the current can make stationary states unstable and induce magnetization oscillation, the feedback can still have important consequences. For instance, we show that the feedback can reduce the linewidth of oscillations, in agreement with experimental observations. A key aspect of this reduction is the suppression of the excitation of short wave length spin waves by the non-local feedback. Finally, we consider nonadiabatic electron transport in narrow domain walls. The non-local feedback in these systems leads to a significant renormalization of the effective nonadiabatic spin transfer torque. These examples show that the self-consistent treatment of spin transport and magnetization dynamics is important for understanding the physics of the coupled dynamics and for providing a bridge between the ongoing research fields of current-induced magnetization dynamics and the newly emerging fields of magnetization-dynamics-induced generation of charge and spin currents.

**Contents**

# 1. Introduction

When electrons flow through systems that include a ferromagnetic region, the flowing electrons become partially spin polarized due to the exchange interaction between conduction electron spins and local magnetizations. Spin transfer torques [1-4] then occur when the spin polarized current passes through another region with a magnetization non-collinear to that in the first region. The spin-polarized current exerts a torque on the non-collinear magnetization by transferring its transverse spin-angular momentum. Spin transfer torques generate a wide variety of magnetization dynamics such as full reversal of magnetization [5,6], steady-state precession [7-10], domain wall motion [11,12], and modification of spin waves [13,14]. These types of current-induced magnetization dynamics could potentially find use in novel functional spintronic devices [15] such as magnetic random access memories (MRAMs) [16], microwave oscillators [17,18], domain wall storage devices [19], and spin wave logic devices [20].

In order to investigate current-induced magnetic excitation, it is essential to formulate the equation of motion of the magnetization affected by spin transport. To do so, spin transfer torques are added to the Landau-Lifshitz-Gilbert (LLG) equation

$$\frac{\partial \mathbf{m}^i}{\partial t} = -\gamma\, \mu_0 \mathbf{m}^i \times \mathbf{H}^i_{\mathrm{eff}} + \alpha \mathbf{m}^i \times \frac{\partial \mathbf{m}^i}{\partial t} + \mathbf{N}^i_{\mathrm{ST}}, \tag{1}$$

where $\mathbf{m}^i$ is the unit vector of the $i^{\mathrm{th}}$ local magnetization, $\mu_0 \mathbf{H}^i_{\mathrm{eff}}$ is the effective magnetic field acting on $\mathbf{m}^i$ (it includes exchange, magnetostatic interactions, anisotropy, thermal fluctuation, and external fields), $\gamma$ is the gyromagnetic ratio of the ferromagnet, $\gamma = (g\mu_B/\hbar)$, $g$ is the Landé g-factor, $\alpha$ is the Gilbert damping constant, and $\mathbf{N}^i_{\mathrm{ST}}$ describes the spin transfer torque acting on $\mathbf{m}^i$. For multilayers such as metallic spin valves or magnetic tunnel junctions where the current flows perpendicular to the interfaces, $\mathbf{N}^i_{\text{ST}}$ is taken as [21,22]

$$\mathbf{N}^i_{\text{ST}} = \gamma[a_J \mathbf{m}^i \times (\mathbf{m}^i \times \mathbf{p}) + b_J (\mathbf{m}^i \times \mathbf{p})], \qquad (2)$$

where $a_J$ and $b_J$ are the coefficients of the in-plane and out-of-plane spin transfer torques, respectively, where the plane is defined to contain two vectors, $\mathbf{m}$ and $\mathbf{p}$, and $\mathbf{p}$ is the direction vector of the pinned-layer magnetization, which is usually assumed to be fixed. On the other hand, when the current flows within a magnetic layer (or nanowire) with a continuously varying magnetization, e.g. domain walls and spin waves, $\mathbf{N}^i_{\text{ST}}$ for one-dimensional system is taken as [23-25]

$$\mathbf{N}^i_{\text{ST}} = u_0 \left( \frac{\partial \mathbf{m}}{\partial x} \bigg|_i \right) - \beta u_0 \left[ \mathbf{m}^i \times \left( \frac{\partial \mathbf{m}}{\partial x} \bigg|_i \right) \right], \qquad (3)$$

where $u_0$ ($= g\mu_B j_e P / 2eM_S$) is the spin current velocity corresponding to adiabatic spin transfer torque, $P$ is the spin polarization, $j_e$ is the charge current density, $M_S$ is the saturation magnetization, and $\beta$ is the ratio of the nonadiabatic spin transfer torque to the adiabatic one [24,25].

Equations (2)-(3) for the spin transfer torque are based on the assumptions that the spin transfer torque depends on the magnetization only instantaneously and locally. Using the instantaneity assumption, $\mathbf{N}^i_{\text{ST}}$ is derived by solving the spin transport equation for relevant systems with fixed (= time-independent) magnetization profiles and then applied to the magnetization dynamics. This instantaneity assumption depends on the ability to decouple the spin transport dynamics from the magnetization dynamics. The decoupling is justified based on the difference in time scales [21,26]. Two characteristic time scales for the spin transport are the spin-flip relaxation time $\tau_{\text{sf}}$ and the spin precession time $h/J$ where $h$ is the Planck constant and $J$ is the interaction energy between conduction electron spins and local magnetizations. Both time scales are of the order of picoseconds or less. On the other hand, the characteristic time scale of the magnetization dynamics is the inverse of the Larmor frequency, which is typically of the order of nanoseconds. Because of this difference in time scales, the two dynamic equations do decouple and the instantaneity assumption is well justified. One can assume that the magnetizations do not vary with time, solve the spin transport equation in the long-time limit to get $\mathbf{N}^i_{ST}$, and feed the result into the equation of motion for the magnetization.

The local approximation is that in Eqs. (2) and (3), $\mathbf{N}^i_{ST}$ is determined by the local values of magnetization (= $\mathbf{m}^i$) and/or local spatial derivative of the magnetization (= $\partial \mathbf{m}/\partial x|_i$). However, the local approximation is not always valid. For example, consider a system consisting of a single ferromagnet (FM) layer sandwiched by two normal metal (NM) layers, where the charge current flows perpendicular to the FM|NM interfaces. The current through the layers generates a spin accumulation, which in turn can generate a spin transfer torque whenever it is not collinear with the magnetization at an interface. Although the direction and magnitude of the spin transfer torque at a point on an interface depends locally on the spin accumulation at the same point, the spin accumulation has an inherently non-local dependence on the magnetization due to spin diffusion. Strictly speaking the spin transfer torque remains local even in this case, but a local interaction between the spin accumulation and the magnetization leads to non-local effective interactions for the magnetization after spin accumulation has been integrated out. In this paper, we call this feedback non-local spin transfer torque because there is a non-local *effective* relation between the spin torque and the magnetization profile. For the torque acting on the single FM layer, lateral spin diffusion in the two neighboring NM layers [27,28] is an important source for non-locality of the torque. Even when net charge flow is perpendicular to the layers, spin diffusion occurs not only along the perpendicular direction but also along the lateral direction (or in-plane direction).

Due to this lateral spin diffusion, spin accumulation at a point in the FM|NM interface depends on the magnetization at other points on the interface within the reach of the spin diffusion. Whenever the magnetization is inhomogeneous in the film plane, the non-local torques will be non-zero. Even if the magnetization is initially in a single domain state, the conventional local spin transfer torques or thermal fluctuations make the magnetization inhomogeneous [29-34] and the non-local torques then becomes non-zero. This non-local spin transfer torque acts as a source of feedback from the magnetization to the spin transport, which, in turn, further affects the magnetization dynamics.

A complete understanding of current-induced magnetic excitations requires a careful treatment of this non-local feedback. In this review, we do so by *self-consistently* solving the two dynamic equations simultaneously, one for magnetization and the other for spin accumulation. In Secs. 2 and 3, we present examples where the self-consistent calculation is essential to capture properties of the coupled dynamics. Section 2 presents the effect of lateral spin diffusion on the magnetization dynamics in layered structures. We first analyze in detail current-induced excitation of a single FM and then the current-driven magnetization oscillation in spin valves that contain two FM layers. Section 3 presents current-induced motion of a narrow domain wall. Here, we use a semiclassical approach to calculate spin transfer torques in the ballistic limit. We end the paper by remarking on the prospects for future work on self-consistent calculation of spin transport and magnetization dynamics.

## 2. Non-local spin transfer torque in layered structures

We consider two types of non-local spin transfer torques in layered structures. One is caused by lateral spin diffusion along the interface of FM|NM. The other is related to the coupling of local spin accumulation along the vertical (thickness) direction of the layers, which is effective when there are more than three ferromagnetic layers. In this section, we focus on the former and briefly discuss the latter in Section 2.4.

*2.1. Basic concept of non-local spin transfer torque due to lateral spin diffusion*

Spin transfer torques caused by lateral spin diffusion, which we will refer to as "lateral spin transfer torque", were proposed by Polianski and Brouwer [27]. The geometry of the system under consideration is shown in Fig. 1: a ferromagnet (FM) of thickness $t_F$ is sandwiched by diffusive non-magnetic layers, $NM_1$ and $NM_2$, of thicknesses $L_1$ and $L_2$, respectively. $NM_1$ and $NM_2$ are connected to reservoirs and the magnetization in the ferromagnet is inhomogeneous. When electrons flow from right to left (charge current *j* flowing from left to the right) the spin filter effect causes minority electrons to accumulate to the right of the ferromagnetic layer and majority electrons to accumulate on the left. Majority electrons have moments parallel to the magnetization but spins that are antiparallel. This difference requires some care in determining the sign of the spin transfer torques.

The two bottom panels in Fig. 1 describe the processes of spin transfer via lateral spin diffusion in detail. On the side of the interface $NM_1$|FM (bottom left panel), after passing through a local magnetization **m**$_1$, a conduction electron spin **s**$_1$ has its moment parallel to **m**$_1$. This **s**$_1$ laterally diffuses along the interface, hits the interface at another point with magnetization **m**$_3$, and then scatters from the interface, transmitting with some probability and reflecting with some probability. The moment of the reflected $s_1$ is anti-parallel to $m_3$ and that of the transmitted electron is parallel to $m_3$. Since the spin angular momentum of $s_1$ changes due to this scattering process, the amount of the change should be transferred to $m_3$ to satisfy the conservation of the spin angular momentum. As a result, $m_3$ experiences spin transfer torque $\tau_1$ that pushes $m_3$ to align with $m_1$; i.e., spin transfer effect on the side of the interface $NM_1|FM$ where the majority spins accumulate tends to suppress any inhomogeneity in the ferromagnetic magnetization. On the other hand, on the side of the interface $FM|NM_2$ (bottom right panel) where minority spins tend to accumulate, the conduction electron spin $s_2$ scattered by a local magnetization $m_1$ initially has its moment anti-parallel to $m_1$. Through the lateral diffusion and the backscattering process by $m_3$, the moment of $s_2$ becomes anti-parallel to $m_3$. This backscattering process generates spin transfer torque $\tau_2$ whose direction is opposite to $\tau_1$; i.e., spin transfer effect on the side of the interface $FM|NM_2$ where the spin accumulation is negative tends to enhance inhomogeneity in the magnetization. Note that the lateral spin transfer torque is inherently non-local because the magnetization everywhere couples together through lateral spin diffusion.

In symmetric systems, $\tau_1$ and $\tau_2$ cancel each other and the lateral spin transfer torque has no net effect. Here we assume that the FM layer is sufficiently thin that the magnetization is uniform along the thickness direction. Making the thickness of $NM_1$ and $NM_2$ different; i.e., $L_1 \neq L_2$, breaks the symmetry and removes this cancellation. The spin accumulation at the interfaces $NM_1|FM$ and $FM|NM_2$ can be found by solving the two second-order differential equations proposed by Valet and Fert [35]. It is straightforward to use Valet-Fert theory in one dimension to show that asymmetric devices give asymmetric spin accumulation.

$$\frac{\partial^2 \mu_S}{\partial z^2} = \frac{\mu_S}{l_{sf}^2} \tag{4}$$

$$\frac{\partial^2 \mu_e}{\partial z^2} = 0 \tag{5}$$

where $l_{sf}$ is the spin diffusion length, $\mu_e$ is the electrochemical potential for the electron density, and $\mu_S$ is the spin chemical potential (that is proportional to the spin accumulation $n_S$ through the Einstein relation $\sigma = e^2 D \partial n / \partial \mu$, where $\sigma$ is the electrical conductivity, $D$ is the diffusion constant, and $n$ is the number density corresponding to the spin accumulation). Figure 2 shows the profiles of $\mu_S$ along the z-axis for symmetric ($L_1 = L_2$, Fig. 2(a)) and asymmetric ($L_1 < L_2$, Fig. 2(b)) structures. We use the boundary condition $\mu_S = 0$ at both interfaces between the non-magnetic layers and the reservoirs. This choice is motivated by the idea that the reservoirs have an infinite density of states. That drives the spin accumulation to zero. Alternatively, placing NM layers at the interfaces with a large spin-orbit coupling, such as Pt or Pt-alloy, induces rapid spin-flip scattering, which also drives the spin accumulation to zero. For a symmetric structure (Fig. 2(a)), the spin accumulations at the left and right interfaces of the FM are of the same magnitude but with the opposite sign, whereas for an asymmetric structure (Fig. 2(b)), they are of different sign and magnitude. Note that Fig. 2(b) describes the case of charge current flowing from $NM_1$ to $NM_2$, where the sum of the spin accumulations at the interfaces of FM|NM is negative. In this case, $\tau_2$ dominates over $\tau_1$; i.e., lateral spin transfer torque tends to increase any inhomogeneities in the magnetization. Reversing the current polarity reverses the spin accumulation so that $\tau_1$ dominates over $\tau_2$; i.e., lateral spin transfer torques suppress inhomogeneities.

*2.2. Previous studies on non-local spin transfer torque due to lateral spin diffusion*

Besides Ref. [27], several experimental [36-38] and theoretical [28,39-41] studies have been performed to understand the lateral spin diffusion effect. Özyilmaz et al. [36]

experimentally observed current-induced excitation of a single ferromagnetic layer. For an asymmetric Cu/Co/Cu nanopillar structure, current-induced excitations were observed for only one polarity of the current, where, according to the prediction [27], the lateral spin transfer torque should increase the magnetization inhomogeneity. In addition, they did not observe such excitations in a symmetric structure, as expected from the discussion above. Özyilmaz et al. [37] also reported experimental results indicating that strong asymmetries in the spin accumulation cause spin wave instabilities in spin valve structures at high current densities, similar to those observed for single magnetic layer junctions.

One of us [28] theoretically extended the initial calculation [27] of lateral spin transfer torque to general situations to allow for variation of the magnetization in the direction of the current-flow. Such variation can give instabilities at a single interface, a possible explanation for spin transfer effects seen in point contact experiments [38]. Brataas et al. [39] reported a theoretical study on the mode dependence of current-induced magnetic excitations in spin valves, and found agreement with the experimental results of Ref. [37]. These calculations [27,28,39] are limited to the linear regime. Even though they identify the onset of instabilities, they do not address the behavior of instabilities after the initial nucleation. Adam et al. [40] performed finite-amplitude self-consistent calculations of spin transport and magnetization dynamics for current-induced magnetic excitations of a thin ferromagnetic layer with asymmetric non-magnetic layers. Their work provided an important proof-of-principle for lateral spin transfer torque, but lacked the spatial resolution and sophistication of full-scale micromagnetic simulations. Hoefer et al. [41] performed a numerical study based on semiclassical spin diffusion theory for a single-layer nanocontact using a convolution approach to calculate the steady-state spin accumulation. They found that directionally controllable collimated spin wave beams can be excited by the interplay of the Oersted field and the orientation of an applied field. These self-consistent calculations [40,41] computed the spin accumulation with either one-dimensional or two-dimensional steady-state solutions of the spin accumulation.

In this section, we show numerical results based on the three-dimensional dynamic solutions of the spin accumulation self-consistently coupled with the magnetization dynamics. Such self-consistent treatments are essential to correctly describe the finite amplitude evolution of the spin wave modes excited by lateral spin transfer torque. They explain two important experimental results: spin wave instabilities in a single FM [36] (Section 2.3.2) and linewidths of precessional oscillations in spin-valves that are narrower than expected from local calculations of the magnetization dynamics [42] (Section 2.3.3).

## 2.3. Self-consistent calculation in layered structures

### 2.3.1. Modeling scheme

We self-consistently solve the equations of motion of local magnetization (Eq. (6)) and spin accumulation $\mathbf{n}_S$ (Eq. (7)) [27,28,39]

$$\frac{\partial \mathbf{m}}{\partial t} = -\gamma_{FM} \mathbf{m} \times \mu_0 \mathbf{H}_{eff} + \alpha \mathbf{m} \times \frac{\partial \mathbf{m}}{\partial t} + \frac{\gamma_{FM}}{M_S t_F} \left[ \mathbf{J}_{S,z} \big|_{z=-t_F/2} - \mathbf{J}_{S,z} \big|_{z=+t_F/2} \right] \quad (6)$$

$$\frac{\partial \mathbf{n}_S}{\partial t} + \sum_{\nu=x,y,z} \nabla_\nu \cdot \mathbf{J}_{S,\nu} = -\gamma_{NM} \mathbf{n}_S \times \mu_0 \mathbf{H}_{ext} - \frac{\mathbf{n}_S}{\tau_{sf}} \quad (7)$$

where $\mathbf{m}$ is the unit vector of local magnetization, $\gamma$ is the gyromagnetic ratio, $\mu_0 \mathbf{H}_{eff}$ is the effective field (including magnetostatic fields, crystalline anisotropy, exchange, current-induced Oersted fields, thermal fluctuations, and external fields ($\mu_0 \mathbf{H}_{ext}$)), $\alpha$ is the intrinsic damping constant, $M_S$ is the saturation magnetization, $t_F$ is the thickness of ferromagnetic layer, $\mathbf{J}_{S,\nu} = -D\nabla_\nu \mathbf{n}_S$ is the spin (number) current density flowing in $\nu$ direction ($\nu = x, y, z$), $D$ is the diffusion coefficient, $\tau_{sf} = l_{sf}^2/D$ is the spin-flip scattering time, and $l_{sf}$ is the spin diffusion length. The change of charge and spin currents ($J_e$ and $\mathbf{J}_S$) at the interface of FM|NM are related to the potential drop over the interfaces as [43,44]

$$J_e(\pm t_F/2) = (G_\uparrow + G_\downarrow)\Delta\mu_e/e + (G_\uparrow - G_\downarrow)\mathbf{m}\cdot(\Delta\boldsymbol{\mu}_S/e) \tag{8}$$

$$\mathbf{J}_S(\pm t_F/2) = -(\hbar/2e^2)\left[(G_\uparrow + G_\downarrow)\mathbf{m}\cdot\Delta\boldsymbol{\mu}_S + (G_\uparrow - G_\downarrow)\Delta\mu_e\right]\mathbf{m} \\ + \mathrm{Re}(G_{\uparrow\downarrow})(\hbar/2e^2)(2\Delta\boldsymbol{\mu}_S\times\mathbf{m} \pm \hbar\partial\mathbf{m}/\partial t)\times\mathbf{m} \tag{9}$$

where $\mu_e$ is the electric potential, $\Delta\mu = \mu(\pm t_F/2 + 0) - \mu(\pm t_F/2 - 0)$ is the potential drop over the interface, $G_s$ (s = ↑ or ↓) is the spin-dependent interface conductance, and the last term proportional to $\partial\mathbf{m}/\partial t$ of Eq. (9) gives the spin-pumping contribution [44], which couples the magnetization dynamics and the spin current. It is characterized by the mixing conductance $G_{\uparrow\downarrow}$. Generally, the mixing conductance has a real and an imaginary part, which couple to the in-plane and out-of-plane terms in the dynamics respectively. Although the out-of-plane spin transfer torque is important in magnetic tunnel junctions [45-53], it is negligible in fully metallic multilayers [54,55]. Thus we neglect Im($G_{\uparrow\downarrow}$) and the associated out-of-plane spin transfer torque. At the interface FM|NM, $J_e$ and $\mathbf{J}_S\cdot\mathbf{m}$ are continuous under the condition of $\boldsymbol{\mu}_S\times\mathbf{m} = 0$ in the ferromagnet. $\boldsymbol{\mu}_S$ and $\mathbf{m}$ are related through Eqs. (7)-(9), and the spin-version of Ohm's law with boundary conditions of $\mu_e = -eV$ (0) and $\boldsymbol{\mu}_S = \mathbf{0}$ (**0**) at the far-right (-left) end of the non-magnetic electrodes. We note that the Eq. (9) is valid for a ferromagnet thinner than the exchange length but thicker than the transverse penetration length.

Since the spin accumulation in NM should be taken into account, the patterned part of Cu leads or spacer is also included in the simulation. Thus, an additional boundary condition for the spin accumulation is required at the side wall of the nano-pillar. We assume that there is no spin-current flow out of the system, i.e., $\partial \mathbf{n}_S / \partial \mathbf{r}_n = 0$, where $\mathbf{r}_n$ is the surface normal vector at the side wall. All simulations repeat two alternating steps: (i) solve Eq. (6) with all boundary conditions to obtain a converged magnetization configuration, and then (ii) solve Eq. (7) to obtain the equilibrium spin accumulation configuration. These steps are repeated. The choice of boundary conditions at the side wall of the nanopillar gives different results than the convolution method used in Ref. [41]. We show that this difference is not important and discuss other differences between the two approaches in Appendix A.

*2.3.2. Single ferromagnet*

In this section, we show the main features of current-induced single ferromagnetic layer excitations, obtained from self-consistent calculations. The layer structure is $Cu_1$ (10 nm) | Co ($t_{Co}$ nm) | $Cu_2$ (52 nm - $t_{Co}$) where $t_{Co}$ varies from 2 nm to 8 nm. As explained above, asymmetric Cu leads provide asymmetric spin chemical potential $\mu_S$ at each side of Co layer. The average spin chemical potential $\bar{\mu}$ at interfaces (= $\mu_S^{Cu1|Co} + \mu_S^{Co|Cu2}$) is negative when the electron flows from the thick to thin Cu layers, corresponding to a negative current. This negative $\bar{\mu}$ provides negative lateral spin transfer torques.

We use the following geometric and magnetic parameters for the single ferromagnetic layer of Co. We consider a nanopillar with an elliptical shape of 60 nm × 30 nm, $M_S$ is 1420 kA/m, the exchange stiffness constant is $2 \times 10^{-11}$ J/m, the gyromagnetic ratio of the ferromagnet and non-magnet are $1.9 \times 10^{11}$ $T^{-1}s^{-1}$ and $1.76 \times 10^{11}$ $T^{-1}s^{-1}$ respectively, we assume there is no anisotropy field, $\alpha$ is 0.01, the unit cell size is 3 nm, and the discretization thickness of the Cu layers varies depending on the total Cu thickness and is not larger than 5 nm. Our results with these cell sizes are converged based on test calculations for a few configurations using smaller cells. The transport parameters for Cu, Co, Pt, and their interfaces are summarized in Table 1. The non-local self-consistent calculation of the dynamics takes approximately 300 times longer than a local calculation.

We calculate magnetic excitations as a function of the out-of-plane field ($\mu_0 H$ = 0 T to 4.6 T) and current ($I$ = –15 mA to +15 mA) at 0 K. Initial magnetic configurations are obtained with applying the out-of-plane field for each case at zero current and zero temperature, and then a current is applied. Figure 3(a) shows the time-averaged out-of-plane component of the magnetization (= $<M_z>/M_S$) as a function of the out-of-plane field and the current for $t_{Co}$ = 2 nm. For positive currents, the magnetization saturates along the out-of-plane direction when the external field $\mu_0 H$ exceeds the out-of-plane demagnetization field $\mu_0 H_d$ ($\approx$1.6 T). However, the magnetization does not saturate at large negative currents even though $H$ is larger than $H_d$, consistent with the data in Ref. [36]. The normalized modulus of the magnetic moment (= $|\mathbf{M}|/M_S$) is smaller than 1 for those bias conditions (Fig. 3(b)), indicating that the magnetic state deviates considerably from the single domain state.

Figure 4 shows snap shots of the magnetization (Fig. 4(a)) and the spin accumulation profiles (Fig. 4(b)-(d)) at $\mu_0 H$ = 2 T and $I$ = –5 mA ($t_{Co}$ = 2 nm). The spin accumulation at the FM|NM interface approximately follows –$\mathbf{M}$ (Fig. 4(b)), whereas the spin accumulation inside of the Cu layer deviates significantly from the local magnetization pattern (Fig. 4(c) and 4(d)) because of spin diffusion. The effect of spin diffusion on the spin accumulation is also seen in Fig. 4(e). The out-of-plane component of averaged spin chemical potential (= $\mu_z$) follows $A\exp(+z/l_{sf}^{Cu})+B\exp(-z/l_{sf}^{Cu})$ where A and B are constants, whereas the in-plane component of averaged spin chemical potential (= $\mu_{xy}$) rapidly decays with increasing the distance from the interface because the spins are mixed during the diffusion process. The decay constant in this case is determined by the characteristic wave vector of the spatial variation, i.e. $l^2 = l_{sf}^2/(1+k^2l_{sf}^2)$, where $k$ is the wave vector characterizing the spatial variation [28].

Figure 5(a) shows color plots of the microwave power for various thicknesses of the Co layer. The microwave power is obtained from the Fourier transformation of the time evolution of $<M_z>/M_S$ where $<...>$ means spatial average. The microwave power is non-zero for the bias conditions where $|\mathbf{M}|/M_S$ is smaller than 1, indicating that the magnetizations are not in stationary states at those bias conditions. The critical current $I_C$ for magnetic excitations depends linearly on $H$ (Fig. 5(a)). It is worthwhile comparing the $I_C$ values obtained from self-consistent calculation with those derived theoretically in the linear limit, which is given by [27,28,39]

$$I_C = \frac{2e}{\hbar} SM_S^2 t_{Co} J_{ex} \frac{\tilde{\alpha}}{S_1}\left(q^2 + \mu_0 \frac{H - H_d}{2J_{ex}M_S}\right). \tag{10}$$

Here $J_{ex}$ is the spin stiffness, $S$ is the area of free layer, $\mu_0 H_d$ is the out-of-plane demagnetization field, and $\tilde{\alpha}$ is the renormalized Gilbert damping constant,

$$\tilde{\alpha} = \alpha + \frac{\gamma_{Co}\hbar^2 \mathrm{Re}(G_{\uparrow\downarrow})}{2M_S t_{Co} e^2} \sum_{\pm} \frac{G_{\pm}(q)}{\mathrm{Re}(G_{\uparrow\downarrow}) + G_{\pm}(q)} \tag{11}$$

where $q$ is the wave number of spin wave, and $G_{\pm}(q)$ is given by

$$G_{\pm}(q) = \frac{\sigma_{Cu}}{2}\sqrt{l_{sf,Cu}^{-2} + q^2}\coth\left(L_{\pm}\sqrt{l_{sf,Cu}^{-2} + q^2}\right) \tag{12}$$

where $\pm$ reads the left and right (or top and bottom) Cu layer.

$S_1$ is the magnitude of the lateral spin transfer torque in dimensionless units,

$$S_1 = \frac{\sigma_{Cu}\mathrm{Re}(G_{\uparrow\downarrow})}{g_m l_{sf,Cu}}\left|\sum_{\pm}\frac{\pm(G_{\pm}(0) - G_{\pm}(q))}{G_{\pm}(0)(\mathrm{Re}(G_{\uparrow\downarrow}) + G_{\pm}(q))}\right|, \tag{13}$$

where $g_m$ is given by

$$g_m = \frac{(G_\uparrow + G_\downarrow)\sigma_{Cu}/l_{sf,Cu} + 2G_\uparrow G_\downarrow \sum_{\pm}\tanh(L_\pm/l_{sf})}{G_\uparrow - G_\downarrow}. \qquad (14)$$

In Fig. 5(b), we compare the calculation results of the slope (= $dI_C/d\mu_0H$) with those obtained from Eq. (10) for various $q$ values. Here, we use the same spin transport parameters as those used in the self-consistent calculations to get the theoretical slopes. The simulation results are in reasonable agreement with analytic ones for $q = \pi/(60$ nm). This good agreement is obtained only when the spin pumping term in Eq. (9) is included. Note that 60 nm is the length of the device along the in-plane easy axis. It suggests that the wavelength of the lowest energy spin wave mode is twice of the device length, due to the geometry and the Oersted field. However, the slopes from the simulations and the analytic results do not agree well with those observed in the experiment (black solid symbols in Fig. 5(b)). This discrepancy may be due to differences between the spin transport parameters used here and the true experimental values.

One aspect of the comparison between theory and experiment that improves going from the analytic model to the full solution is the intercept of the extrapolated boundary at $I = 0$. From Eq. (10), the theoretical intercept at $I = 0$ is the out-of-plane demagnetization field $\mu_0H_d$. The value of $\mu_0H_d$ slightly decreases from 1.6 T to 1.4 T as $t_{Co}$ increases from 2 nm to 8 nm, caused by the change in the demagnetizing factors depending on the geometry of FM junction. In the experiment of Ref. [36], however, the intercept is found to be much smaller than $\mu_0H_d$. The simulated results of the intercept are also considerably smaller than $\mu_0H_d$, and the intercept decreases with increasing $t_{Co}$, as shown in the inset of Fig. 5(b). Thus, the intercepts obtained from the self-consistent calculation are in better agreement with the experimental observations than the theoretical ones. We attribute this better agreement to the fact that the self-consistent model more realistically takes into account the influence of the shape and finite size of the nano-pillar on the spin wave mode as we discuss below.

Figure 6(a) shows the time evolution of $<M_z>/M_S$ for various negative currents for $\mu_0 H = 2.5$ T. The magnetization initially saturates along the out-of-plane direction because of the large out-of-plane field. When the current is turned on, a very small in-plane component of the magnetization develops especially at the long edges where the Oersted field is the largest. The interplay between this laterally inhomogeneous magnetization and negative lateral spin transfer torque excites spin waves, resulting in a rapid decrease in $<M_z>$ within a few nanoseconds.

To understand spin wave mode excitation by lateral spin transfer torques, we perform an eigenmode analysis for the magnetization dynamics (Fig. 6(b)-(d)). To calculate eigenmodes, we choose the bias condition of $I = -11$ mA and $\mu_0 H = 2.5$ T, which shows a periodic oscillation of $<M_z>$. We note that such periodic oscillations are observed only for some bias conditions and the magnetic excitation is highly nonlinear in general. The spectral density of $<M_z>$ shows two peaks at two frequencies, $f_L$ ($\approx$ 75 GHz) and $2f_L$ ($\approx$ 150 GHz) (Fig. 6(b)), where $f_L$ satisfies $f_L = \gamma_{Co} \mu_0 H/2\pi$. On the other hand, for a single domain state, we expect the precession frequency to be $\gamma_{Co} \mu_0 (H - H_d <M_z>/M_S)/2\pi$ because the effective magnetic field experienced by the magnetizations is the summation of the external field and the internal demagnetization field. At $I = -11$ mA, $<M_z>/M_S$ is about 0.6 (Fig. 6(a)); in this approximation, the precession frequency would be 46 GHz, which is much smaller than the obtained precession frequency $f_L$. This disagreement indicates that the precession frequency is mostly determined by the external out-of-plane field $\mu_0 H$, and that contributions from $\mu_0 H_d$ are negligible. An eigenmode analysis of the spatial patterns (Fig. 6(c) and 6(d)) for the two peak frequencies gives some insight into this peculiar field dependence. The eigenmode images are obtained from local power spectrum $S_z(\mathbf{r}, f)$ [57]

$$S_z(\mathbf{r},f) = \left| \sum_j M_z(\mathbf{r},t_j) \exp(i 2\pi f t_j) \right|^2 . \tag{15}$$

The precession region with a higher power is localized at the edges. These eigenmodes are unique features originating from lateral spin transfer torque and not expected for the field-driven excitation [57]. Figures 6(e)-6(h) show the time evolution of the magnetic domain patterns at the same bias condition. The magnetization near the edges is mostly in the plane, but near the center of the cell, it is in vortex-like states. The peculiar frequency dependence on the field could be explained by the formation of vortex-like states where the demagnetization field along the thickness direction is significantly reduced.

### 2.3.3. Spin valves

In this section, we apply the self-consistent non-local model to a spin-valve structure with two ferromagnetic layers experimentally studied by Sankey et al. [42], Cu(80) | Py(20) | Cu(6) | Py(2) | Cu(2) | Pt (Py = Permalloy), with all thicknesses in nm. They found that the resonances excited by current have narrower linewidths at low temperatures than expected from a finite temperature macrospin calculation. This reduced linewidth indicates that some additional effect can improve the coherence time of precession in nanomagnets.

We use the same parameters for Cu as used in the previous section and replace the parameters for Co by parameters for Py provided by the Cornell group. The pillar has an elliptical shape with 120 nm × 60 nm, $M_S$ is 645 kA/m [42], the exchange stiffness constant is $1.3\times10^{-11}$ J/m, $\alpha$ is 0.025 [10], the unit cell size is 5 nm. The transport parameters for Py and Py|Cu are summarized in Table 1. We assume the magnetization of the pinned layer (Py 20 nm) is fixed along the in-plane easy axis and that it gives no stray field. While the pinned layer is likely not to be fixed in reality, we keep it fixed to focus on the effect of lateral diffusion. For finite temperature simulations, we add the Gaussian-distributed random fluctuation field [58] (mean = 0, standard deviation = $2\alpha k_B T/(\gamma M_S V \Delta t)$, where $\Delta t$ is the integration time step, $V$ is the volume of unit cell) to the effective field for magnetization. We test convergence of the stochastic calculations and find that the results are converged for $\Delta t$ below 50 fs based on the average magnetization along the easy axis. For stochastic simulation, one may require temperature- and cell-size-dependent renormalization of parameters in order to take into account effect of magnons having a shorter wavelength than the unit cell size employed in simulations. Several ways to renormalize the exchange constant and the saturation magnetization have been proposed [59,60]. However, we are not aware of any way to renormalize the damping constant and the spin transfer torques. These parameters are of critical importance for the calculation of current-induced magnetic excitations. In this work, we do not consider temperature- and cell-size-dependent renormalization of parameters. We also neglect any temperature dependence of the transport parameters.

To investigate whether or not the reduced linewidth originates from lateral spin transfer torques, we perform numerical simulations based on three different approaches: i) a macrospin model (MACRO), ii) a conventional micromagnetic model without considering lateral spin transfer torque (CONV), and iii) a non-local, self-consistent model (SELF). Fig. 7 shows contours of spectral density of $<M_X>$ as a function of the current at the temperature $T=$ 4 K when a field of 50 mT is applied along the in-plane easy axis (// $x$). Positive current corresponds to the electron-flow from Cu(2) to Cu(6), and thus positive lateral spin transfer torque. The macrospin simulations show the familiar red- and blue-shift depending on the bias current $I$ (Fig. 7(a)). The conventional simulations show only a red-shift up to a critical current ($I_C^{CONV} \approx 2$ mA, Fig. 7(b)). Here, the magnetization dynamics becomes complicated due to excitation of incoherent spin-waves when $I > I_C^{CONV}$. As indicated by an arrow, secondary peaks are observed at about half of the frequency of main peaks, corresponding to the precession of end domains [31]. In the non-local, self-consistent simulations, similar secondary peaks are observed, indicating deviations from a single domain state, but peak structures are much clearer than they are in the conventional simulations up to about 2.4 mA, which is larger than $I_C^{CONV}$ (Fig. 7(c)). The blue-shift followed by a transition region is also observed. It indicates that positive lateral spin transfer torques improve the coherence of the magnetization dynamics.

Figure 8(a) shows the power spectra computed in the three models ($I = 1.4$ mA and $T = 10$ K). It is evident that at low temperature, the non-local, self-consistent simulations give the narrowest linewidth. We calculate the temperature ($T$) dependence of linewidth from Lorentzian fits (Fig. 8(b)). At low temperatures ($T < 50$ K), the non-local, self-consistent simulations provide narrower linewidths than other two approaches, consistent with experimental observations [42]. However, we observe that the linewidths computed from the macrospin simulation are wider than those computed from the conventional micromagnetic simulation. This counterintuitive result may be due to the fact that the linewidth is affected by the precession angle [8]. By estimating the precession angle of micromagnetic results from the spatial average of magnetization component, we find that the macrospin and micromagnetic models give different precession angles whereas two micromagnetic models give similar angles at the bias current. Because of these limitations, direct comparisons of the linewidths between the macrospin and micromagnetic simulations may be limited. Below, we discuss effect of the self-consistent feedback on the linewidth by comparing the two micromagnetic modeling approaches; this comparison would be relatively free from the above-mentioned limitations and shows that the feedback reduces the linewidth.

From Fig. 8(b), we find that the non-local, self-consistent model gives a narrower linewidth than the conventional micromagnetic model for $T < 50$ K. It suggests that the coupling among local magnetizations induced by positive lateral spin transfer torque indeed results in a substantial improvement of the coherence time of precession at a low temperature. For $T > 50$ K, however, the linewidth in the non-local self-consistent simulation increases more rapidly than in the conventional micromagnetic simulation. We note that this does not mean that the positive lateral spin torque makes the linewidth very broad at high temperatures. As shown in Fig. 8(c), the more rapid increase in the linewidth in the non-local self-consistent simulations originates from a mode splitting. We find that power spectra calculated from the non-local self-consistent simulations consist of two peaks; a narrow main peak at a higher frequency indicated by up-arrows, and a secondary broad peak at a lower frequency. The frequency of the secondary peak does not change much with temperature, whereas that of the main peak increases slightly with temperature. This kind of mode splitting has been observed in experiments [61] and numerical studies based on a conventional micromagnetic model with no lateral spin torque [62,63]. Because of this mode splitting, the linewidth obtained from the fit using a single Lorentzian function increases rapidly with temperature.

In the low-temperature limit, two nonlinear effects of the positive lateral spin transfer torques may cause the narrower linewidths in the non-local, self-consistent simulations: an increase of the effective exchange stiffness at short range and an increase of the damping of incoherent spin-waves at long range. As a result, positive lateral spin transfer torques provide an additional nonlinearity to the spin-wave damping. For spin-torque nano-oscillators, the linewidth $\Delta\omega$ in the low-temperature limit (i.e. $T$ < 10 K in our case) is given by [64,65]

$$\Delta\omega = \Gamma_+(P_0)\left(\frac{k_B T}{E_0}\right)\left[1+\left(\frac{N}{\Gamma_{eff}}\right)^2\right] \quad (16)$$

where $\Gamma_+(P) = \Gamma(1+QP)$ is the positive damping of the oscillator, $\Gamma$ is the equilibrium linewidth in the passive region, $Q$ is a phenomenological coefficient characterizing the nonlinearity of the positive damping, $P$ is the normalized power, $k_B T$ is the thermal energy, $E_0 = \omega_0 V_0 M_S (I-I_C)/[\gamma\mu_0(I+QI_C)]$ is the average energy of the stable auto-oscillation, $\omega_0$ is the ferromagnetic resonance frequency, $I_C$ is the critical current for the magnetic excitation, $N = d\omega(P)/dP$ is the nonlinear frequency shift coefficient obtained from $\omega(P) = \omega_0 + NP$, $\Gamma_{eff} = \sigma_I(I+QI_C)$ is the effective nonlinear damping, $I$ is the bias current, $\sigma_I = \varepsilon g\mu_B/2eMV_0$, $\varepsilon$ is the spin-polarization efficiency, $V_0$ is the volume, $P_0 = (\zeta-1)/(\zeta+Q)$ is the equilibrium oscillation power, and $\zeta = I/I_C$ is the supercriticality.

Equation (16) predicts two important consequences of the nonlinearity. First, the linewidth of an auto-oscillator with a nonlinear frequency shift (i.e. $N \neq 0$) increases by the factor $(1+(N/\Gamma_{eff})^2)$ from that of a linear oscillator (i.e. $N = 0$). Second, the linewidth of a nonlinear oscillator decreases with increasing nonlinear damping $Q$ when $N$ is large. The linewidth is determined by nonlinear properties of the system where the normal linear damping is compensated by local spin transfer torques. In this case, an increase of the nonlinear damping can lead to a decrease of the linewidth, known as noise suppression due to nonlinear feedback [66,67] which has been widely observed in various fields such as optics [68], mechanics [69], and biology [70]. While this nonlinear feedback typically requires an external feedback element, in spin-valves it is inherent.

Figure 8(d) shows that *N* is evidently nonzero and almost identical in both the conventional micromagnetic simulations and the non-local, self-consistent simulations. Thus, in both approaches, the linewidth increases from that expected for a linear oscillator. Using equation (16), we fitted the values of *Q* from the calculated linewidths at $T$ = 10 K and obtained *Q* = 0.12 in the conventional micromagnetic simulations and *Q* = 1.96 in the non-local, self-consistent simulations. The fitted value *Q* in the latter is consistent with the *assumed* values (*Q* = 1 to 3) in the Ref. [65,71] to explain experimental observations. It should be noted that in Ref. [65,71], the large *Q* is purely phenomenological. Our non-local self-consistent treatment suggests that the large *Q* may be caused by the lateral spin diffusion. Thus, the nonlinear spin-wave damping due to lateral spin transfer torque is probably responsible for narrower linewidths in the non-local, self-consistent simulations at low temperatures. For the opposite polarity of the current (i.e. negative lateral spin transfer torque), we observe an increase of the linewidth (not shown) as would be expected for the case when lateral spin diffusion enhances inhomogeneity.

*2.4. Summary*

To summarize this section, we report non-local, self-consistent calculations for current-induced excitation of a single ferromagnetic layer and spin valves. The former are in good agreement with previous theoretical [27,39] and experimental studies [36]. They provide an improved understanding of the coupled dynamics between magnetizations and spin transport, and the excitation of spin wave modes for negative lateral spin transfer torques. In case of a single ferromagnetic layer, only a negative net lateral spin transfer torques lead to spin wave instabilities, while positive net lateral spin transfer torques do not. In spin valve structures, self-consistent calculations are crucial for correct evaluation of the oscillation linewidth.

Whereas the conventional spin transfer torque and its interplay with the Oersted field tend to cause a large amplitude incoherent spin wave excitation [29-32], the positive lateral spin transfer torque effect captured by the self-consistent calculation tends to reduce spatial inhomogeneities (suppress spin waves) and leads to more coherent magnetization dynamics at low temperatures. This effect would be beneficial for microwave oscillators utilizing spin transfer torque, where a narrow linewidth is a key requirement.

Lateral spin transfer torques are non-zero when the following three conditions are satisfied: (i) the spin accumulation at the two interfaces of a ferromagnetic layer are asymmetric, (ii) at least one of neighboring layers is diffusive, and (iii) the magnetization is inhomogeneous. Condition (i) is generally satisfied in multilayer structures (= NM | FM (pinned) | NM | FM (free) | NM) since there is a pinned ferromagnet on one side of the free ferromagnet whereas there is no ferromagnet on the other side. Condition (ii) is also generally satisfied for fully metallic multilayers and even for magnetic tunnel junctions. In a typical magnetic tunnel junction, the free ferromagnet is sandwiched by an insulator and a diffusive non-magnet (capping layer). The lateral spin diffusion is allowed only in the capping layer, which maximizes the net lateral spin transfer torque because the lateral spin transfer torque at the other interface is essentially zero. Finally, condition (iii) is almost always satisfied because the current-induced Oersted field is inhomogeneous and leads to inhomogeneous magnetizations in all but the strongest saturating fields [31]. Furthermore, thermal fluctuations of the magnetization are spatially inhomogeneous. Therefore, lateral spin transfer torques are usually non-zero.

Finally, we briefly comment on another type of non-local spin transfer torque in multilayers. Let us consider a spin valve containing three FMs; i.e., $FM_1$ | $NM_1$ | $FM_2$ | $NM_2$ | $FM_3$, where $FM_1$ is pinned (= pinned layer) and other two FMs serve as a synthetic free layer.

Such multilayer structures with a synthetic free layer are of considerable interest for MRAM applications [72-77] and spin transfer torque-oscillators [78,79]. In this structure, not only are there spin transfer torques at the $NM_1|FM_2$ interface, but also at the $FM_2|NM_2$ and $NM_2|FM_3$ interfaces (Fig. 9). Furthermore, the spin transfer torques at each interface depend on the orientation of both of the other magnetizations (Fig. 9(c)), because local spin accumulations at each interface are vertically coupled through the whole layer structure. In this kind of structure, the spin transfer torque is non-local even without the lateral spin diffusion, and requires a self-consistent calculation to investigate current-induced magnetic excitation [80-82].

# 3. Non-local spin transfer torque for a magnetic domain wall

*3.1. Current-induced motion of a domain wall*

A magnetic domain wall is the transition region between two magnetic domains in which the magnetization continuously varies. The interplay between the magnetic exchange on one hand and the crystalline anisotropy and the magnetostatic interactions on the other hand gives a finite width to the wall. An electrical current passing through a domain wall in a ferromagnetic nanowire can move the wall, because the current creates a spin transfer torque. Current-induced domain wall motion has been intensively studied both theoretically and experimentally. Understanding this motion requires understanding the coupling between conduction electron spins and the continuously varying magnetization. It may also find use in storage and logic devices in which the domain wall is used as the information unit (for comprehensive reviews about current-induced domain wall motion based on local spin transfer torque, please see Refs. [83-87]).

*3.2. Non-local spin transfer torque for a narrow domain wall*

One of the central issues for current-induced domain wall motion is how to reduce the threshold current density to move the domain wall. The reason is twofold. A typical threshold current density for a metallic ferromagnet is about $10^{12}$ A/m$^2$ [11,88]. Such high current densities cause significant Joule heating, making it difficult to distinguish spin transfer effects from heating effects [89-93]. From an application point of view, devices need to operate with current densities lower than this threshold current density to minimize electromigration. For this reason, there has been substantial research directed toward reducing the threshold current density. Several solutions have been proposed. One approach is to use resonant dynamics of domain wall motion by controlling current pulse widths [94] or injecting consecutive current pulses [95]. Another approach is to reduce the hard-axis anisotropy that the spin transfer torque must overcome to move a domain wall. Such reductions can be achieved by shaping nanowire geometries properly since the hard-axis anisotropy is caused by geometry-dependent demagnetizing effects, as predicted theoretically [96] and verified experimentally [97].

Yet another approach is to increase the nonadiabatic spin transfer torque, which controls the wall motion for small currents in ideal nanowires. When electrons flow through a spatially slowly varying magnetization configuration, their moments tend to stay aligned with the magnetization. Since this requires the moments to rotate, there must be a reaction torque on whatever is causing them to rotate, i.e. the magnetization. The reaction torque has the form of the first term in Eq. (3) [1,2] and is referred to as the adiabatic spin transfer torque because it comes from the spins "adiabatically" following the magnetization. The other term in Eq. (3), is perpendicular to the adiabatic spin transfer torque and is referred to as the nonadiabatic spin transfer torque even though some contributions to it occur in the adiabatic limit. Without the nonadiabatic torque, the adiabatic torque in combination with the other terms in the LLG equation leads to intrinsic pinning for currents below a threshold [23]. Intrinsic pinning happens because the wall distorts as it moves and the distortion leads to torques that oppose the motion. The nonadiabatic spin transfer torque acts like a magnetic field for domain walls and thus makes the threshold current density vanish for an ideal nanowire. The larger the nonadiabatic torque, the faster the domain wall motion for small currents.

The importance of the nonadiabaticity for current-induced domain wall motion, has led to a number of theoretical [23-25,98-106] and experimental studies [94,107-112] to determine the nonadiabatic spin torque parameter $\beta$. Several mechanisms for the nonadiabatic spin transfer torque have been proposed. One class of mechanisms is based on the changes in processes that contribute to magnetic damping change in the presence of current. These changes typically have the form of the nonadiabatic torque. For example, a phenomenological treatment of the scattering of itinerant electrons by spin-dependent impurities generates both damping and a nonadiabatic spin transfer torque in the presence of current [24]. Similarly, band structures with spin-orbit coupling and electron scattering give both damping [113] and nonadiabatic torques [103], both of which can be calculated from first principles [104,105]. The nonadiabatic torque due to these mechanisms does not depend on the domain wall width. For domain walls much wider than the characteristic length scales of spin transport, these mechanisms are the only ones that make the spin current deviate from the magnetization direction and give a non-adiabatic spin transfer torque.

Additional mechanisms become more significant as domain walls get narrower. For moderately narrow domain walls (width ≈ 5 nm to 10 nm), spin diffusion can increase the effective nonadibaticity [114,115]. For narrower domain walls (width < 5 nm), the conduction electron spins traversing the domain wall cannot follow a sharp change in the magnetization and thus contribute to the nonadiabaticity [100,102]; i.e., ballistic spin-mistracking. When the domain wall is extremely narrow (i.e., one or two atomic layers), momentum transfer can occur due to the reflection of electron spins from the domain wall [23]. This class of mechanisms (spin diffusion, spin mistracking, and momentum transfer) generally gives non-local spin transfer torques and their contributions depend on the domain wall width.

Initial experiments for current-induced domain wall motion in metallic systems have used $Ni_{80}Fe_{20}$ (Permalloy) for which domain wall widths are large (≈ 100 nm). The theoretical predictions for the enhanced nonadiabaticity by reducing the domain wall width have encouraged experimentalists to study systems with smaller domain wall widths by utilizing materials with strong perpendicular anisotropy [110,112]. For narrow domain walls, the role of non-local spin transfer torque on the domain wall motion may be important.

*3.3. Previous studies on self-consistent calculation for current-induced domain wall motion*

Manchon et al. [114] theoretically predicted that the spin diffusion generates an additional spin transfer torque that effectively enhances the nonadiabatic torque. This new torque is inversely proportional to the square of the domain wall width and strongly depends on the domain wall structure. For instance, it can increase the transverse velocity of vortex cores in vortex domain walls, whereas its influence remains negligible for transverse domain walls. This dependence on the domain wall structure arises from the fact that the spin diffusion current transverse to the electron-flow direction is significant for a vortex wall but negligible for a transverse wall. Recently, Claudio-Gonzalez et al. reported numerical results based on a self-consistent calculation of the drift-diffusion model and the LLG equation [115]. They found that an increase in the effective nonadiabaticity for a vortex wall but only minimal changes for a transverse wall, consistent with the theoretical prediction of Ref. [114].

For Bloch or Nèel walls formed in perpendicularly magnetized nanowires, this spin diffusion torque does not enhance the effective nonadiabaticity because the wall is a simple one-dimensional domain wall in contrast to vortex walls. Then, unless the domain wall is extremely narrow, ballistic spin-mistracking will be the important mechanism for changing the nonadiabatic torque. Ohe et al. [116] performed self-consistent calculations to investigate this effect based on a lattice model [117] where the conduction electrons are treated quantum mechanically and thus spin mixing in the states of the conduction electrons is fully taken into account. They found that when the Fermi energy of the electrons is larger than the exchange energy (i.e., a typical situation for transition metals), spin precession induces spin-wave excitations in the local magnetization. This spin-wave excitation contributes to the domain wall displacement at low current densities but reduces the domain wall velocity for large current densities as compared to the adiabatic limit.

Here, we present self-consistent calculations of the non-local spin transfer torque based on a semiclassical, free electron approach. Our approach differs from the previous self-consistent calculation [116] in two aspects. One difference is the determination of which electron states are occupied. In a Landauer picture, the Fermi levels of the leads are fixed and different. The Fermi level of the material between the leads adjusts in response to the applied voltage to create local charge neutrality. This adjustment leads to current flow that is half excess electrons moving forward and half a deficit of electrons moving backwards. Ref. [116] introduced extra right-propagating electrons in the energy range $E_F < E < E_F + eV$ where $V$ is the voltage drop across the nanowire (Fig. 10(a)). Since electrons were added to the equilibrium Fermi sea, charge neutrality was violated in their calculation. In contrast, we induce extra right-propagating electrons in the energy range $E_F < E < E_F + eV/2$ and remove left-propagating electrons in the energy range $E_F - eV/2 < E < E_F$ (Fig. 10(b)), so that charge neutrality is preserved. The difference in occupancy results in an important difference in the spatial distribution of non-local spin transfer torque between Ref. [116] and our work. Figures 10(c) and (d) show the spatial distribution of spin transfer torque $\mathbf{N}_{ST}$ obtained from the two approaches. Here, the spin transfer torque $\mathbf{N}_{ST}$ is separated into two vector components, $\mathbf{N}_{ST} = \mathbf{N}_{ST}^{Adia} + \mathbf{N}_{ST}^{Nonadia}$, where $\mathbf{N}_{ST}^{Adia}$ and $\mathbf{N}_{ST}^{Nonadia}$ are aligned along $\partial\mathbf{m}/\partial x$ and $\mathbf{m} \times \partial\mathbf{m}/\partial x$, respectively. In Ref. [116] the oscillatory non-local spin transfer torque appears at only one side of the domain wall, whereas in our work it appears at both sides of the domain wall (see Fig. 10(c) and (d)). An additional difference between the calculations is that Ref. [116] assumes one-dimensional mesoscopic transport by considering only a single-electron channel (*k*-normal, **k** // *x*), whereas we treat the non-equilibrium spins over the full three-dimensional Fermi surface. Treating the full Fermi surface generates spin dephasing because of the variation of the precession length over the Fermi surface. Figure 10(c) shows that for a spin transfer torque calculation with a single-electron channel of Ref. [116], the non-local oscillation of spin transfer torque is very significant and does not decay even far from the domain wall. In contrast, the oscillation is suppressed at large distances from the wall in our approach due to the strong spin dephasing (Fig. 10(d)).

*3.4. Semiclassical approach*

Here we use a semiclassical approach proposed by one of us [100], which is based on two main approximations; i.e., ballistic transport and a parabolic band structure. With these approximations, we show that mistracking torques can make important contributions to domain wall dynamics. For all but extreme cases, these contributions can be captured through effective values of local parameters. This simple model maximizes the importance of the non-local effects, but since the effects can be largely be accounted for by a local approximation, our use of the "best case" is appropriate. We expect a local parameterization to be even more appropriate when scattering and realistic band structures are taken into account.

Before explaining the model details, let us discuss the relevance of this simple model. We expect the ballistic limit to be appropriate for materials with very short domain wall widths (about 1 nm), which are shorter than the mean free path. A ballistic transport picture becomes less appropriate when domain wall widths are greater than mean free paths and precession lengths. In that case, we expect that scattering will reduce the non-local effects obtained from a ballistic transport model. We also expect that non-local effects will be weaker for realistic band structures than for parabolic band structures because dephasing is stronger for realistic band structures. Thus, we expect that the results for a parabolic band structure set an upper limit for the importance of non-local effects. We show below that in most cases we consider, the non-local effects can be accounted for by suitably renormalized local parameters. We expect that conclusion to be even stronger for more realistic band structures in domain walls in which scattering is important. The Hamiltonian is

$$H = -\frac{\hbar^2}{2m}\nabla^2 - \mu\boldsymbol{\sigma}\cdot\mathbf{B}_{ex}(x), \qquad (17)$$

where $\boldsymbol{\sigma} = (\sigma_x, \sigma_y, \sigma_z)$ is a vector composed of the three Pauli matrices and $\mu = g\mu_B$. $\mathbf{B}_{ex}(x)$ is aligned along the local magnetization everywhere and describes the magnetic field experienced by a conduction electron spin through the s-d exchange coupling. Its magnitude is defined as

$$E_{ex} = 2\mu|\mathbf{B}_{ex}| = \hbar^2 k_B^2 / m. \qquad (18)$$

The Fermi wave vectors for up and down spins, $k_F^+$ and $k_F^-$ are given by

$$k_F^\pm = \sqrt{k_F^2 \pm k_B^2}, \qquad (19)$$

where the Fermi energy is $E_F = \hbar^2 k_F^2 / 2m$. The spatial evolution of the single-particle spin density $\mathbf{s}_\pm(x, k_x)$ for a given energy $E$ is obtained from

$$\frac{d\mathbf{s}_\pm(x, k_x)}{dx} = -\frac{k_B^2}{\langle k \rangle}\mathbf{s}_\pm(x, k_x) \times \hat{\mathbf{B}}_{ex}(x), \qquad (20)$$

where $\langle k \rangle = (k_+ + k_-)/2$, and $k_+$ and $k_-$ for $\mathbf{s}_+(x, k_x)$ are defined by

$$k_\pm^2 = 2mE/\hbar^2 \pm k_B^2 \qquad (21)$$

with $E = \hbar^2(k_x^2 - k_B^2)/2m$, and $k_+$ and $k_-$ for $\mathbf{s}_-(x, k_x)$ are defined by Eq. (21) with $E = \hbar^2(k_x^2 + k_B^2)/2m$. The semiclassical single-electron spin-current density is then obtained from

$$\mathbf{J}_{\pm}(x, k_x) = \mathbf{s}_{\pm}(x, k_x) \frac{\hbar k_x}{m}. \tag{22}$$

One finds the spin current density $\mathbf{J}_{\pm}(x)$ by integrating $\mathbf{J}_{\pm}(x, k_x)$ over the Fermi surface in the presence of an electric field $E\hat{\mathbf{x}}$,

$$\mathbf{J}_{\pm}(x) = \int \left[ f_{\pm}\left(\mathbf{k} - \frac{eE\tau}{\hbar} \hat{\mathbf{x}}\right) - f_{\pm}(\mathbf{k}) \right] \mathbf{J}_{\pm}(x, k_x) \frac{d^3\mathbf{k}}{(2\pi)^3}, \tag{23}$$

where the spin-dependent Fermi-Dirac distribution function $f_{\pm}(\mathbf{k}) = \Theta(k_F^{\pm} - |\mathbf{k}|)$ implies that the distribution of electrons outside the region of inhomogeneous magnetization are characteristic of the zero-temperature bulk [100]. Then, $\mathbf{J}(x) = \mathbf{J}_{+}(x) + \mathbf{J}_{-}(x)$ is the total spin current density, and the spin transfer torque is given by

$$\mathbf{N}_{ST}(x) = -\frac{\partial \mathbf{J}(x)}{\partial x}. \tag{24}$$

We plug this semiclassical calculation of spin transfer torque in to the LLG equation, Eq. (1). At every integration time step, we compute the semiclassical calculation of the non-local spin transfer torque for a given magnetization profile, and then update the magnetization dynamics using the spin transfer torque for the next time step. This procedure is repeated and as a result, the effect of spin transfer torques on the magnetization dynamics and subsequent feedback are taken into account self-consistently.

Several remarks on the computation are in order. First, the length of the nanowire treated in the calculation should be much longer than the domain wall width. If not, unphysical nonequilibrium spin density can arise from discontinuities at the edges. Second, multiscale modeling is important to reduce the computation time. In this work, the unit cell size for calculating the LLG equation is more than 10 times larger than for calculating the semiclassical spin transport equation. The smaller cell size for the spin transport calculation is essential to ensure a convergence when solving Eq. (20).

*3.5. Current-induced domain wall motion by non-local spin transfer torque*

A qualitative explanation for the origin of the non-local spin transfer torque is as follows. When the domain wall is sufficiently wide compared to the precession period of the spin density determined by $k_F$ and $k_B$, the precession amplitude of the spin density is small and averaged out when integrated over the Fermi surface. As a result, the local spin direction of spin current is almost perfectly aligned along the local magnetization direction, so that spin transfer torque can be locally defined by the gradient of the local magnetization. In contrast, when the domain wall is narrow and its width is comparable to the precession period, the precession amplitude is considerable even at points far from the domain wall and the spin transfer torque becomes non-local.

In this work, we carry out micromagnetic simulations for a semi-one dimensional nanowire (i.e., the nanowire is discretized along the length direction, but not along the width or the thickness direction), self-consistently coupled with a semiclassical spin transport calculation. We assume a perpendicularly magnetized nanowire with the following parameters: the Fermi energy $E_F = 10$ eV, the exchange splitting $E_B = 1$ eV, the exchange constant $A_{ex} = 1 \times 10^{-11}$ J/m, the saturation magnetization $M_S = 1300$ kA/m, the Gilbert damping constant $\alpha = 0.03$, the nanowire width = 200 nm, and the nanowire thickness = 4 nm. The perpendicular crystalline anisotropy constant $K_u$ is varied from $2 \times 10^6$ J/m$^3$ to $1 \times 10^7$ J/m$^3$ in order to vary the domain wall width $\lambda_{DW}$. The local nonadiabaticity ($\beta$) caused by the spin relaxation is assumed to be zero in order to focus on the non-local spin transfer torque caused by the ballistic spin-mistracking.

Figures 11(a) and (b) show three vector components of spin transfer torque for $K_u = 2 \times 10^6$ J/m$^3$ ($\lambda_{DW} \approx 2.71$ nm) and $K_u = 10^7$ J/m$^3$ ($\lambda_{DW} \approx 0.98$ nm), respectively. The non-locality of the spin transfer torque becomes more pronounced for a smaller $\lambda_{DW}$; i.e., the amplitude of oscillatory spin transfer torque is larger, and the non-zero spin transfer torque is observed further away from the domain wall. Figures 11(c) and (d) show $\mathbf{N}_{ST}^{Adia}$ and $\mathbf{N}_{ST}^{Nonadia}$ for various $\lambda_{DW}$ values, respectively. Two observations are worth noting. First, the non-local nonadiabatic contribution of spin transfer torque (Fig. 11(d)) becomes more significant as $\lambda_{DW}$ gets smaller. Second, both $\mathbf{N}_{ST}^{Adia}$ and $\mathbf{N}_{ST}^{Nonadia}$ are non-local (Fig. 11(c)).

Figure 12 shows the domain wall velocity $v_{DW}$ as a function of the spin current velocity $u_0$, obtained from the self-consistent calculation. We did not observe any significant spin wave excitations, in contrast to Ref. [116]. We attribute this difference to the fact that the non-local spin transfer torque is not as significant as in Ref. [116] due to the strong spin dephasing (see Fig. 10). Overall trends of $v_{DW}$ are similar to those expected from the local approximation with nonzero local nonadiabaticity $\beta$ [24,25]. When the spin current velocity $u_0$ (proportional to the current density) is small, $v_{DW}$ is linearly proportional to $u_0$, and the slope $v_{DW}/u_0$ in the linear range increases with decreasing $\lambda_{DW}$. When $u_0$ exceeds a certain threshold ($u_{WB}$, indicated by down arrows in Fig. 12), $v_{DW}$ deviates from the linear dependence. The threshold $u_{WB}$ corresponds to the Walker breakdown [83,118], above which the domain wall undergoes a precessional motion. These overall trends of $v_{DW}$ indicate that the non-local spin transfer torque indeed acts like an additional local nonadiabatic spin transfer torque.

We can understand the similarity of the domain wall motion from a collective coordinates approach to analyze the calculation results obtained from the self-consistent model. Following Thiele's work [119], we assume the domain wall structure is $\mathbf{m} = (\sin\phi\sin\theta, \cos\phi\sin\theta, \cos\theta)$ where $\sin\theta = \mathrm{sech}[(x-X(t))/\lambda_{DW}]$, $\cos\theta = \tanh[(x-X(t))/\lambda_{DW}]$, and $\phi = \phi(t)$. Here, $X$ is the domain wall position, $\phi$ is the domain wall tilt angle, $\lambda_{DW}$ is the domain wall width, and $t$ is time. After some algebra, one obtains the equations of motion of the collective coordinates ($X$, $\phi$) in the rigid domain wall limit (i.e., $\partial\lambda_{DW}/\partial t = 0$),

$$\frac{\partial \phi}{\partial t} + \frac{\alpha}{\lambda_{DW}}\frac{\partial X}{\partial t} = \frac{\tilde{c}_J}{\lambda_{DW}} \tag{25}$$

$$-\frac{1}{\lambda_{DW}}\frac{\partial X}{\partial t} + \alpha \frac{\partial \phi}{\partial t} = -\frac{\tilde{b}_J}{\lambda_{DW}} - \gamma \frac{K_d}{M_S}\sin 2\phi \tag{26}$$

$$\tilde{c}_J \equiv -\frac{\lambda_{DW}}{2}\int_{x=-\infty}^{x=+\infty} dx \left[ \left(\mathbf{m}\times \mathbf{N}_{ST}^{Nonadia}\right)\cdot\frac{\partial \mathbf{m}}{\partial X}\right] \tag{27}$$

$$\tilde{b}_J \equiv \frac{1}{2}\int_{x=-\infty}^{x=+\infty} dx \left[ \left(\mathbf{m}\times \mathbf{N}_{ST}^{Adia}\right)\cdot\frac{\partial \mathbf{m}}{\partial \phi}\right] \tag{28}$$

where $K_d$ is the hard-axis anisotropy of domain wall. In the local approximation, one can recover $\tilde{b}_J \equiv u_0$ and $\tilde{c}_J \equiv \beta u_0$ using $\mathbf{N}_{ST}^{Adia} = u_0 \partial \mathbf{m}/\partial x$ and $\mathbf{N}_{ST}^{Nonadia} = -\beta u_0 (\mathbf{m}\times \partial \mathbf{m}/\partial x)$. In our case, however, $\tilde{b}_J$ and $\tilde{c}_J$ can be obtained by integrating Eqs. (27) and (28) numerically, because of the non-local nature of both $\mathbf{N}_{ST}^{Adia}$ and $\mathbf{N}_{ST}^{Nonadia}$. We define $\eta_{\text{eff}}$ ($\equiv \tilde{b}_J/u_0$) and $\beta_{\text{eff}}$ ($\equiv \tilde{c}_J/u_0$) that effectively describe the average adiabaticity ($\approx$ effective spin polarization) and nonadiabaticity of non-local spin transfer torque, respectively. The dependence of $\eta_{\text{eff}}$ and $\beta_{\text{eff}}$ on $\lambda_{DW}$ are summarized in Fig. 13. $\eta_{\text{eff}}$ is close to 1 for a large $\lambda_{DW}$ and decreases with decreasing $\lambda_{DW}$. In contrast, $\beta_{\text{eff}}$ is close to 0 for a large $\lambda_{DW}$ and increases with decreasing $\lambda_{DW}$. The changes in $\beta_{\text{eff}}$ are much more significant than those in $\eta_{\text{eff}}$. Given the uncertainty in the proper parameters to describe these systems, it is likely that change in $\eta_{\text{eff}}$ will be much more difficult to observe than those in $\beta_{\text{eff}}$.

Based on Eqs. (25) to (28), one can define several important physical quantities of domain wall dynamics (see Appendix B for details). The threshold spin current velocity $u_{WB}$ for the Walker breakdown, the domain wall velocity $v_{\text{steady}}$ for $u_0 < u_{\text{WB}}$, and the average domain wall velocity $\bar{v}$ for $u_0 \gg u_{\text{WB}}$ are given by

$$u_{\text{WB}} = \frac{\gamma K_d \lambda_{\text{DW}}}{M_S} \frac{\alpha}{|\beta_{\text{eff}} - \eta_{\text{eff}} \alpha|}, \tag{29}$$

$$v_{\text{steady}} = \frac{\beta_{\text{eff}}}{\alpha} u_0, \tag{30}$$

$$\bar{v} = \frac{\eta_{\text{eff}} + \alpha \beta_{\text{eff}}}{1 + \alpha^2} u_0. \tag{31}$$

In Fig. 14, we show how well this local approximation for $\eta_{\text{eff}}$ and $\beta_{\text{eff}}$ shown in Fig. 13 can describe the self-consistent calculation results shown in Fig. 12. When they agree, there is no need for the full self-consistent solution. Instead, one can calculate $\eta_{\text{eff}}$ and $\beta_{\text{eff}}$ based on the semiclassical calculation in Eqs. (27) and (28), and use them in the LLG equation with the local approximation for spin transfer torque. When it is valid, this procedure significantly reduces the computation time. The plots of $v_{\text{DW}}$ versus $u_0$ are mostly similar in the two approaches (Fig. 14(a)-(e)), but there are some discrepancies. An important discrepancy is the Walker breakdown threshold, $u_{WB}$. For instance, when $K_u$ is $3 \times 10^6$ J/m$^3$ (the equilibrium $\lambda_{\text{DW}} \approx 2.03$ nm) (Fig. 14(b)), $u_{\text{WB}}$ for the self-consistent calculation is about 310 m/s whereas $u_{\text{WB}}$ for the local approximation is about 220 m/s. This difference in $u_{\text{WB}}$ is caused by the fact that $\lambda_{\text{DW}}$ changes in the simulation but is treated as a constant in deriving the local approximation. As the current increases, the domain wall tilt angle $\phi$ also increases. This change in $\phi$ causes a change in $K_d$ and in turn, a change in $\lambda_{\text{DW}}$. Figure 14(f) shows $\lambda_{\text{DW}}$ in the steady state ($\lambda_{\text{DW}}^{\text{Steady}} \equiv \lambda_{\text{DW}}$ at $t \to \infty$) versus $u_0$ for $K_u = 3 \times 10^6$ J/m$^3$. $\lambda_{\text{DW}}^{\text{Steady}}$ for a small $u_0$ is close to its equilibrium value (= 2.03 nm), but decreases with increasing $u_0$. As shown in Fig. 13, the reduced domain wall width results in an increased $\beta_{\text{eff}}$; in this case, $\beta_{\text{eff}} \approx$ 0.019 for $u_0$ = 5 m/s whereas $\beta_{\text{eff}} \approx 0.022$ for $u_0$ = 300 m/s. Using these values of $\beta_{\text{eff}}$ to Eq. (29) with $\eta_{\text{eff}} \approx 1$ and $\alpha$ = 0.03, one finds that $u_{\text{WB}}$ indeed changes substantially due to this nonlinear effect as shown Fig. 14(b). We conclude that the local approximation with $\eta_{\text{eff}}$ and $\beta_{\text{eff}}$ calculated from spin transport equations would capture the core effect of the non-local spin transfer torques qualitatively, but it cannot reproduce the results obtained from the self-consistent calculation quantitatively unless they are artificially adjusted.

*3.6. Summary*

To summarize this section, we show self-consistent calculations for current-induced dynamics of narrow domain walls. We find that for narrow domain walls, the self-consistent calculations predict the spin transfer torque to be non-local and spatially oscillatory due to the ballistic spin-mistracking mechanism. The non-local spin transfer torque generates domain wall motion and thus its effect is similar to the local nonadiabatic spin transfer torque. However some of its effect such as the Walker breakdown threshold value cannot be fully captured by the local nonadiabatic spin transfer torque approximation. Therefore when $\lambda_{\text{DW}}$ is close to 1 nm, it is necessary to adopt the self-consistent calculations for quantitative description of current-driven domain wall motion. It is worth comparing our result to available experimental ones. Thomas et al. [94], Heyne et al. [109], and Eltschka et al. [111] have found that vortex cores exhibit a much larger nonadiabaticity ($\beta \approx 8\alpha$ to $10\alpha$) compared to transverse domain walls ($\beta \approx \alpha$). According to our result, this large nonadiabaticity of vortex cores is unlikely to be caused by the ballistic spin-mistracking since a typical width of a vortex core is about 10 nm. The large reported values of $\beta$ in these systems are more likely to be related to spin diffusion effect [114,115] and/or anomalous Hall effect [120]. On the other hand, Burrowes et al. [112] have tested very narrow Bloch-type domain walls of about 1 nm using FePt nanowires and found that such a narrow domain wall does not cause a significant increase in the nonadiabaticity. This experimental result is inconsistent with our self-consistent calculation. Assuming that $\lambda_{DW}$ in the experiment is indeed around 1 nm, there are a few possible reasons for this discrepancy. Our model assumes a spherical Fermi surface with the free-electron approximation. However, the shape of a realistic Fermi surface usually deviates substantially from a sphere. If a realistic Fermi surface was considered, the contribution from non-local spin transfer torque is likely to be reduced because of additional spin dephasing due to the complicated Fermi surfaces as we mention earlier in Sec. 3.4. Another possible reason for the inconsistency is that the experiment of Ref. [112] used a thermally activated depinning from a point defect to estimate the nonadiabaticity. Since the width of FePt nanowires in the experiment is about 200 nm, it is reasonable to assume that a domain wall could bend when escaping from a point defect. If this is the case, our one-dimensional model calculation should not be compared to this experiment since a two-dimensional domain wall structure may cause an additional spin dephasing. Therefore, we believe that better defined measurements should be done to experimentally test the role of the non-local spin transfer torque due to ballistic spin-mistracking for narrow domain walls.

Although there are some ambiguities in directly comparing our model calculation to experiments, our result indicates that it may be important to perform self-consistent calculations to understand current-induced dynamics of narrow domain walls in detail. Since many recent experiments have utilized materials systems with high perpendicular magnetic anisotropy, combining experimental measurements and self-consistent calculations would be essential to understand the underlying physics and to design efficient domain wall devices.

# 4. Conclusion and outlook

In this review, we present self-consistent calculations of transport and magnetization dynamics for several representative examples. The self-consistent treatment allows us to capture the core effect of the feedback from the magnetization dynamics to the spin transport and back to the magnetization dynamics through non-local spin transfer torques. The feedback results in current-induced excitation of a single ferromagnetic layer, a narrower linewidth for magnetization oscillation in spin valves, and an additional effective nonadiabatic spin transfer torque for domain wall dynamics. These examples show the importance of self-consistent treatments of spin transport and magnetization dynamics for understanding the physics of the coupled dynamics.

Before ending this review, we remark that the examples discussed so far are not the only cases for which a self-consistent treatment is required. In the following, we will briefly comment on other examples where the feedback mechanism is non-trivial.

Giant magnetoresistance is often considered as an inverse effect of spin transfer torque. However, the generation of spin currents by magnetization dynamics would more aptly be considered the inverse process since the spin transfer torque is the excitation of magnetic dynamics by spin currents. These processes, which generate spin currents by magnetic dynamics, are spin pumping [44,121,122] and the spin motive force [123-125]. Spin currents cannot be directly measured, but they couple to other processes that can. In ferromagnets, spin currents generate charge currents, which in turn generate electric voltages [126-135], and the generation of spin currents enhances magnetic damping [136-142].

Just as spin transfer torques in multilayers and nanowires are similar processes in different geometries, so are spin pumping and the spin motive force. Spin pumping occurs in bilayer structures where a ferromagnetic layer is attached to a non-magnetic layer [44,121,122]. A precessing magnetization in the ferromagnet pumps a spin current into the non-magnet transferring energy and angular momentum from the ferromagnet to the conduction electrons of the non-magnet. This transfer increases the magnetic damping rate in the ferromagnet. However, the pumped spin current generates a spin accumulation in the non-magnet. This spin accumulation in turn generates a back-flow current back into the ferromagnet through diffusion processes. The quantitative enhancement of the Gilbert damping [44] and the voltage drop across the interface [126] requires proper treatment of the balance between the pumping and back-flow currents. One approach for such calculations is the magnetoelectronic circuit theory used in Section 2.

The spin motive force, on the other hand, is found in systems with a single ferromagnet [123-125] like a magnetic nanowire. When the magnetization varies in both space and time, conduction electrons experience a spin-dependent electric field that generates spin and charge currents. Early calculations of the spin motive force [123-125,128-130] and the consequent enhancement of Gilbert damping [136-141] did not consider other processes that might be important: spin accumulation, spin diffusion, and spin-flip scattering. However, just as it is necessary to properly consider the backflow current for a description of spin pumping, so is it necessary to consider these processes for a calculation of the spin motive force. Several of us have investigated these effects theoretically, and found that spin relaxation processes [142] significantly modify the spin motive force. For example, charge currents are perfectly canceled by diffusion currents in one-dimensional systems. Spin currents become non-local and become smaller depending on the characteristic length of spatial variation of the magnetization and the spin diffusion length. For such one-dimensional systems, we provided an analytical expression of spin motive force including spin relaxation processes [142]. For

two- or three-dimensional systems, however, such analytical solutions are not available so that self-consistent calculations would be necessary to describe the coupled dynamics.

Self-consistent calculations would also be very important for descriptions of spin transfer torques and spin motive forces in ferromagnetic systems with strong spin-orbit coupling, for example, ferromagnets with Rashba interactions. Obata and Tatara [143], and Manchon and Zhang [144] independently predicted the existence of field-like spin transfer torque induced by in-plane current in Rashba ferromagnets. A number of experimental [145-150] and theoretical [151-159] studies have followed this work. Miron et al. reported that an in-plane current-induced field-like spin torque is present for Pt|Co|AlO$_x$ structures where the inversion symmetry is broken [145]. Miron et al. also reported that a domain wall in such structures moves against the electron-flow direction with high speed [146]. This reversed domain wall motion with high speed cannot be explained by conventional adiabatic and nonadiabatic spin transfer torques, but may be explained by a damping-like spin transfer torque in addition to all other spin transfer torques (i.e., adiabatic, nonadiabatic, and the field-like torques) [156] and the Dzyaloshinskii-Moriya interaction [159]. The damping-like spin transfer torque may originate from a spin Hall effect in a heavy metal layer like Pt [159-165] and/or a nonadiabatic correction to the field-like torque [155-158]. This damping-like torque also allows switching the magnetization by in-plane currents [149,164,166].

At present, the appropriate description of this unconventional current-induced magnetization dynamics is still controversial. It is not clear whether an explanation based on the spin Hall effect, Rashba spin-orbit coupling, both, or something else, is appropriate for all experiments or individual experiments. To resolve this controversy, it may be important to develop a model that takes into account both types of spin-orbit effects and computes the properties of spin transfer torques accurately. For instance, we have developed a Boltzmann transport model considering the two sources of spin transfer torques (i.e., the spin Hall effect and Rashba spin-orbit coupling) and found that both sources can generate not only field-like torques but also damping-like torques for thin ferromagnets [165]. In a different approach, we have found [167] that for two-dimensional electron gases and under the assumption that the spin-orbit potential is comparable to the exchange interaction, the field-like spin torque has a complicated dependence on the angle between the current direction and the magnetization direction. In this case, self-consistent calculations are needed to properly take into account the effect of complicated angle-dependent spin transfer torque on current-induced magnetization dynamics. Furthermore, since spin transfer torques and spin motive forces are closely related, a sizable spin transfer torque due to Rashba spin-orbit coupling suggests that the magnetization dynamics in Rashba ferromagnets can generate a large spin motive force [168-170]. In this case, the spin motive force may require self-consistent calculations to accurately account for the spin relaxation process since the Rashba spin-orbit coupling correlates the spin directions with the wave vectors.

Up until this point, we have discussed the coupled dynamics of charge, spin, and magnetization. Another important degree of freedom is heat. Temperature gradients across structures may also generate spin transfer torques just as voltage gradients do [171-178]. Recently, the existence of thermal spin transfer torques was experimentally demonstrated in metallic spin valves [175] and theoretically predicted in magnetic tunnel junctions [178]. This type of torque mediated by magnon- and/or spin-wave-spin current may find use in moving domain walls [179-185]. It is closely related to spin-dependent thermoelectric effects, such as spin-dependent Seebeck, Peltier, and Nernst effects [186-189]. These heat- and spin-dependent phenomena are unexplored largely at the moment and thus would require various self-consistent calculations that couple heat, spin, and magnetization dynamics all together.

**Appendix A. Comparison of the convolution method to a full solution of the spin accumulation profiles in the lateral spin diffusion problem**

Ref. [41] introduced a convolution method that leads to the speed up in the calculation of the lateral diffusion. Since the speed gain is substantial, it is important to test the validity of the underlying approximations. Here we do so by examining our full solutions of the drift-diffusion equation.

In the convolution method, the spin chemical potential $\mu_S$ at a point **r** is given by $\boldsymbol{\mu}_S(\mathbf{r}) = \int dv\, \widetilde{\mathbf{K}}(\mathbf{r}-\mathbf{r}')\mathbf{m}(\mathbf{r}')$, where the kernel $\widetilde{\mathbf{K}}(\mathbf{r}-\mathbf{r}')$ is a 3 by 3 tensor that relates $\mu_S$ at **r** to the magnetization **m** at a different point **r**′. Its explicit form is given in Ref. [41]. In the convolution method, the kernel $\widetilde{\mathbf{K}}$ is assumed to depend on (**r**−**r**′) but not explicitly on **r** itself. This assumption leads to substantial speed up in the computing time because the kernel can be precomputed and the convolution can be done with fast numerical techniques. Several approximations underlie this approach. It assumes that the kernel does not change near boundaries in the structure and assumes that the magnetization only has small deviations from the average magnetization.

Here we test the errors that are introduced by the convolution method in nanopillars in which all of the layers have been patterned. Figure A1(a) shows a schematic of a system consisting of NM (10 nm) | FM (8 nm) | NM (32 nm). The layers have been patterned into a nanopillar 41 nm wide, and the spin diffusion lengths are chosen to be 200 nm in the non-magnet and 10 nm in the ferromagnet. Other parameters are similar to those of Py/Cu in the main text. Arrows in the ferromagnet show local magnetizations. The magnetization points in the x direction except for the cell located at x = $x_0$ where it is in the z direction. Figure A1(b) shows the z-component of spin chemical potential $\mu_z$, calculated by our approach, in the NM region at the bottom interface of FM|NM for two cases; $x_0 = 0$ (center of nanopillar) and $x_0 = 18$ nm (close to an edge). In case of $x_0 = 0$, the spin chemical potential profile is symmetric along the lateral direction (i.e., x-axis) whereas it is slightly asymmetric due to the boundary effect in case of $x_0 = 18$ nm as indicated by arrows in Fig. A1(b). However, the two agree surprisingly well. In part, this arises because the spin accumulation is much more local than would be expected from the long spin diffusion length. The spin accumulation is more local because the interface with the ferromagnet and the interface with the reservoir acts as effect spin flip scattering sites. Unless the lead is very thick, the spin diffusion length becomes largely irrelevant compared to the spin flip scattering at the interfaces.

The convolution approach will break down when the magnetization varies significantly compared to its average value. We illustrate this point in a spin-valve structure with domain walls in both layers. The problem with the convolution method used in Ref. [41] for this situation is that the kernel is for the transverse magnetization based on a solution for the longitudinal transport that is uniform across the device. This assumption is clearly violated in the structure considered here with domain walls (see Fig. A1(c)).

Overall, the convolution method is a convenient approximation to calculate the spin accumulation profiles in some cases because it uses significantly less computation time compared to full calculations. However, this method is not reliable in all situations. For instance, it breaks down for calculations of magnetization reversal, particularly when the reversal mode is non-uniform. In contrast, full calculations can be applied to all cases at the cost of time-consuming calculations.

# Appendix B. Collective coordinates approach for non-local spin transfer torque in a narrow domain wall

With $\eta_{\text{eff}}$ and $\beta_{\text{eff}}$, Eqs. (25) and (26) can be rewritten as,

$$\frac{\partial \phi}{\partial t} + \frac{\alpha}{\lambda_{\text{DW}}} \frac{\partial X}{\partial t} = \frac{\beta_{\text{eff}} u_0}{\lambda_{\text{DW}}}, \tag{B.1}$$

$$-\frac{1}{\lambda_{\text{DW}}} \frac{\partial X}{\partial t} + \alpha \frac{\partial \phi}{\partial t} = -\frac{\eta_{\text{eff}} u_0}{\lambda_{\text{DW}}} - \gamma \frac{K_d}{M_S} \sin 2\phi. \tag{B.2}$$

When $u_0$ is smaller than $u_{\text{WB}}$, $\phi$ increases in the initial time stage and then becomes saturated to a certain value over time. In this limit (i.e., $\partial \phi / \partial t = 0$ as $t \to \infty$), we find

$$\sin 2\phi = \frac{M_S}{\gamma \alpha K_d \lambda_{\text{DW}}} (\beta_{\text{eff}} - \eta_{\text{eff}} \alpha) u_0. \tag{B.3}$$

$u_{\text{WB}}$ is determined from the maximum of R.H.S. of Eq. (B.3), since the absolute value $\phi$ of is maximized at $u_0 = u_{\text{WB}}$. Thus, $u_{\text{WB}}$ is given as

$$u_{\text{WB}} = \frac{\gamma K_d \lambda_{\text{DW}}}{M_S} \frac{\alpha}{|\beta_{\text{eff}} - \eta_{\text{eff}} \alpha|}. \tag{B.4}$$

When $\beta_{\text{eff}} = 0$ and $\eta_{\text{eff}} = 1$, Eq. (B.4) recovers the spin current velocity for the Walker breakdown ($u_{\text{WB}} = \gamma K_d \lambda_{\text{DW}} / M_S$) driven by the local adiabatic spin transfer torque [23,190].

When $u_0 < u_{\text{WB}}$, the domain wall moves steadily. In this case, domain wall velocity $v_{\text{steady}}$ is obtained from Eq. (B.1) with $\partial \phi / \partial t = 0$ as

$$v_{\text{steady}} = \frac{\beta_{\text{eff}}}{\alpha} u_0. \tag{B.5}$$

When $u_0 \gg u_{\text{WB}}$, $\partial \phi / \partial t$ is always nonzero and domain wall undergoes a continuous precession motion. In the limit of very large current, one obtains the average velocity $\bar{v}$ by averaging Eqs. (B.1) and (B.2) over a period of the precession of $\phi$ and using $\langle \sin 2\phi \rangle = 0$ where $\langle ... \rangle$ is the time-average over a period;

$$\bar{v} = \frac{\eta_{\text{eff}} + \alpha \beta_{\text{eff}}}{1+\alpha^2} u_0. \tag{B.6}$$

**Acknowledgements**

We thank B. Dieny, A. Vedyayev, A., G.E.W. Bauer, A. Kent, J.Z. Sun, I.N. Krivorotov, A.N. Slavin, J.-V. Kim, S. Zhang, A. Manchon, R.D. McMichael, and T.J. Silva for fruitful discussions. This work was supported by the NRF (2010-0023798, 2011-028163), by the MEST Pioneer Research Center Program (2011-0027905), and KU-KIST School Joint Research Program. K.J.L. acknowledges support under the Cooperative Research Agreement between the University of Maryland and the National Institute of Standards and Technology Center for Nanoscale Science and Technology, Award 70NANB10H193, through the University of Maryland.

Table 1

Transport parameters for numerical simulations shown in section 2.3.2 and 2.3.3.

| | Cu[a] | Co[a] | Permalloy[b] (Py) | Co\|Cu[a] | Cu\|Pt[a] | Py\|Cu[b] |
|---|---|---|---|---|---|---|
| Bulk resistivity $\rho$ (nΩ·m) | 6.0 | 75 | 255 | | | |
| Bulk spin asymmetry, $\beta_s$ | 0 | 0.46 | 0.7 | | | |
| Spin diffusion length $l_{sf}$ (nm) | 450 | 59 | 5.5 | | | |
| Diffusion coefficient $D$ ($\times 10^{15}$ nm$^2$s$^{-1}$) | 41 | 1.7 | 1.7 | | | |
| Interfacial resistance[c] AR* (fΩ·m$^2$) | | | | 0.51 | 0.12 | 0.97 |
| Interfacial spin asymmetry[c], $\gamma_s$ | | | | 0.77 | 0 | 0.77 |
| Spin mixing conductance Re[$G_{\uparrow\downarrow}$] ($\times 10^{14}$ Ω$^{-1}$m$^{-2}$) | | | | 5.5 | - | 6.0 |

[a] Transport parameters for Cu, Co, Pt, and their interfaces are obtained from the literature [54, 56].

[b] Transport parameters for Py are provided by Cornell group.

[c] Spin-dependent conductance $G_s$ (s = $\uparrow$ or $\downarrow$) in Eq. (9) is related with AR* and $\gamma_s$ through $(G_\uparrow + G_\downarrow) = 1/AR^*(1-\gamma_s^2)$ and $(G_\uparrow - G_\downarrow) = \gamma_s/AR^*(1-\gamma_s^2)$.

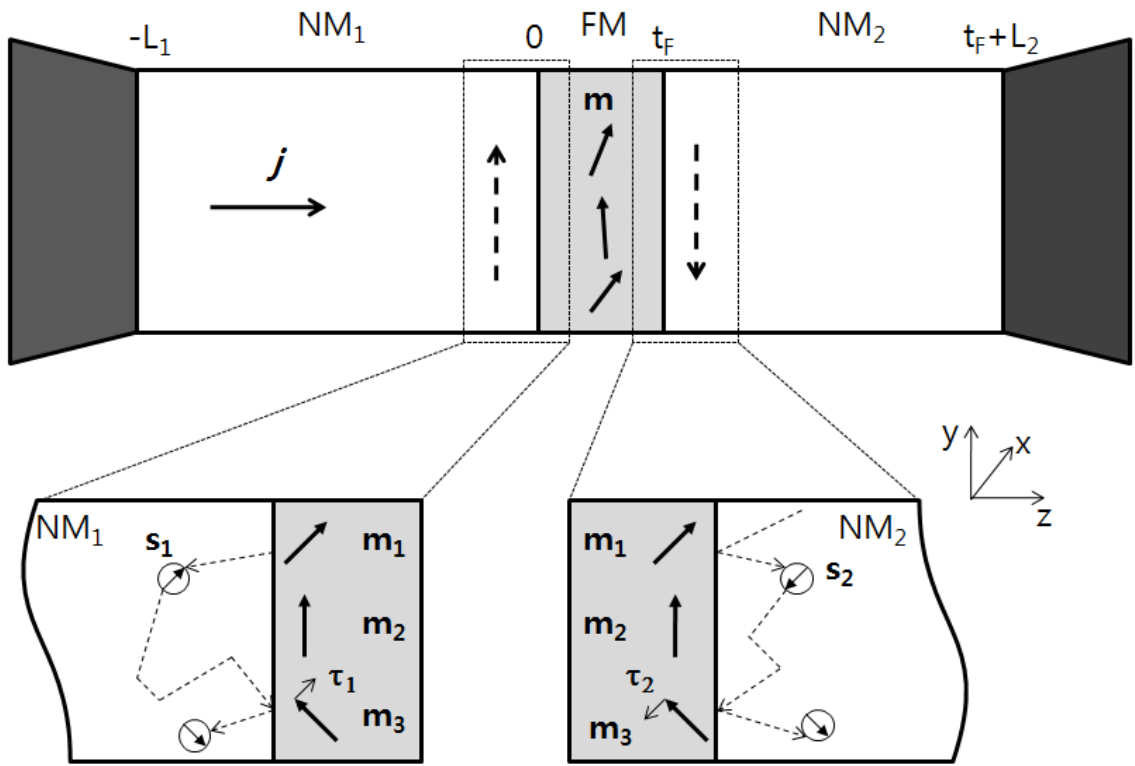

Fig. 1. Top: schematic picture of a nanopillar structure consisting of a normal metal (NM$_1$) | a ferromagnetic layer (FM) | a normal metal (NM$_2$). Bottom: cartoons of lateral spin diffusion at left and right interfaces of FM.

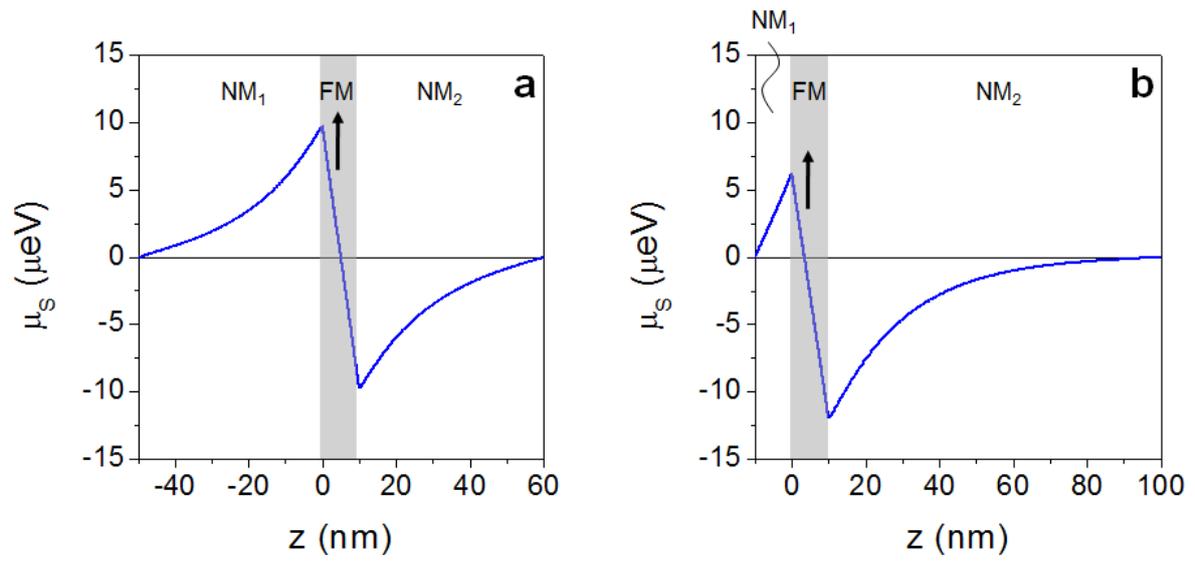

Fig. 2. Spin chemical potential patterns in (a) a symmetric structure and (b) an asymmetric structure, calculated from one-dimensional Valet-Fert theory [35].

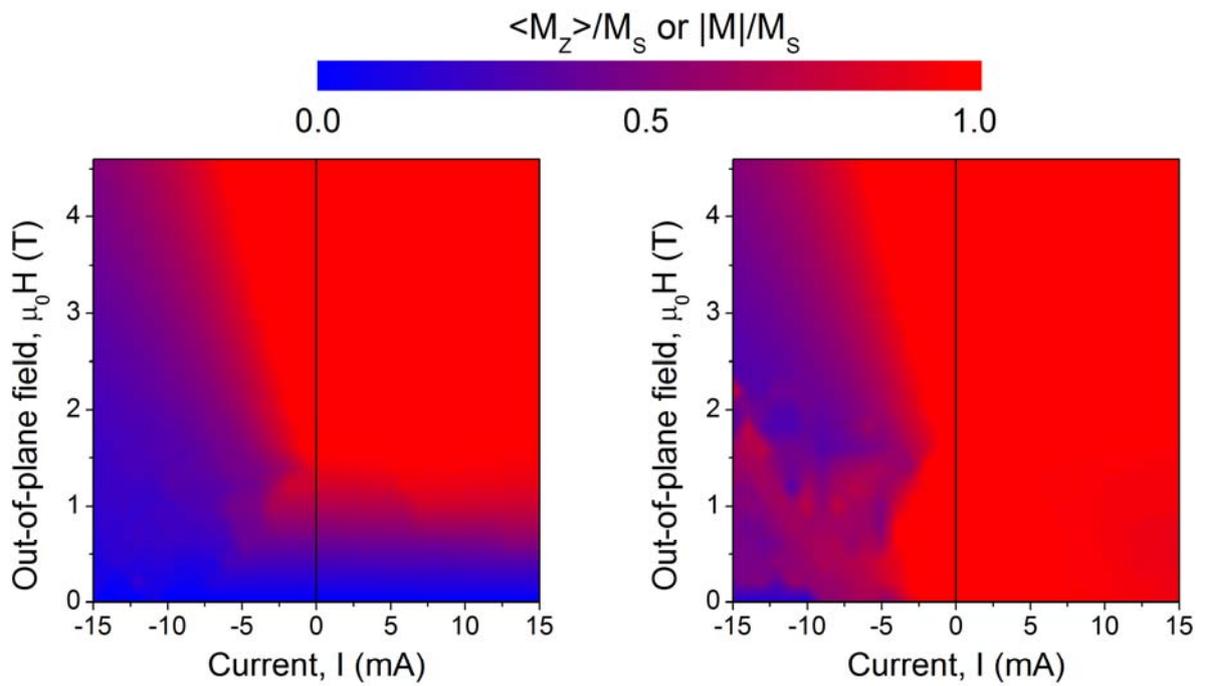

Fig. 3. Current-induced excitation of single ferromagnetic layer sandwiched by asymmetric Cu layers: (a) Out-of-plane magnetization $<M_z>$ as a function of the out-of-plane field and current. (b) Normalized modulus of the magnetic moment as a function of the out-of-plane field and current. 1 mA corresponds to about $7.07\times10^{11}$ A/m$^2$.

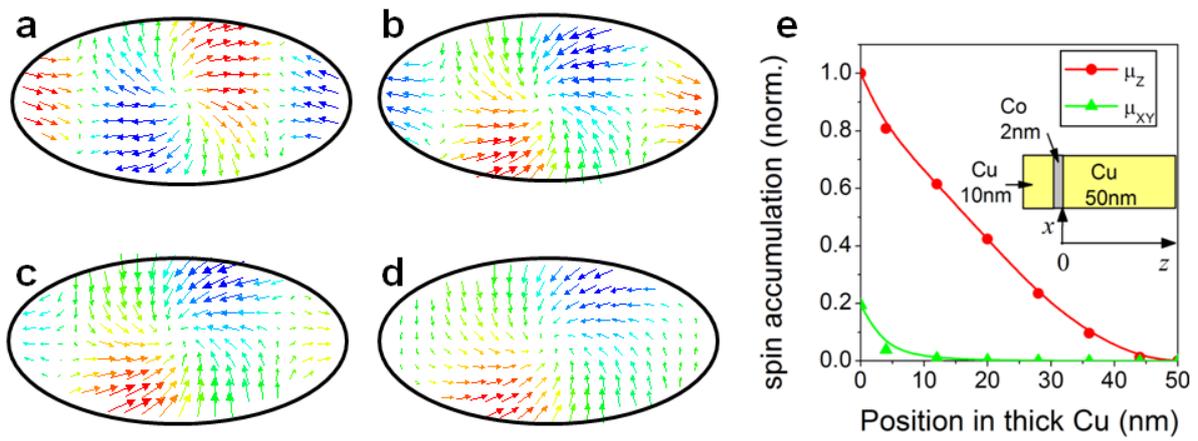

Fig. 4. Vector plots of the magnetization pattern (**M**) and the spin accumulation (**n**$_S$) patterns in a thick Cu layer at $\mu_0 H$ = 2 T and $I$ = -5 mA ($t_{Co}$ = 2 nm). (a) **M** in Co layer. (b) **n**$_S$ (×1) at interface ($z$ = 0 nm). (c) **n**$_S$ (×5) at $z$ = 6 nm. (b) **n**$_S$ (×15) at $z$ = 18 nm. (e) Normalized average in-plane and out-of-plane components of the spin accumulation as a function of the distance from the interface of Co|Cu. In (e), the lines are guides to the eye.

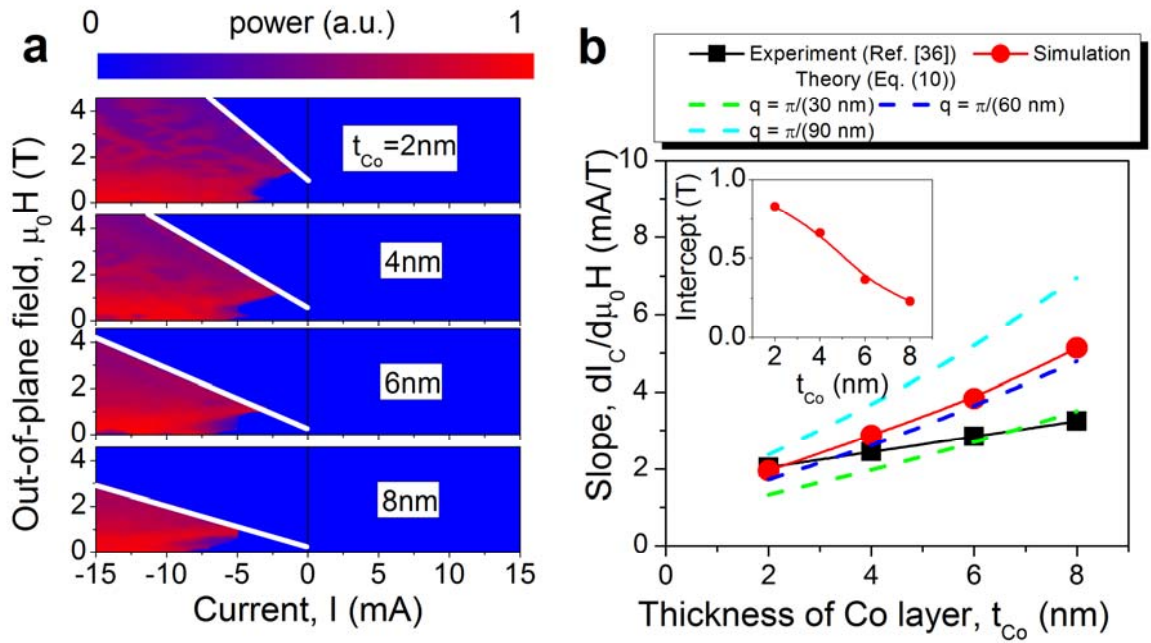

Fig. 5. Dependence of the threshold current for current-induced excitation on the thickness of single ferromagnetic layer: (a) Microwave power at various Co layer thicknesses. White lines correspond to phase boundaries. (b) Slope of the phase boundary as a function of the thickness of Co layer. Inset of (b) shows the intercept of the extrapolated boundary as a function of the thickness of Co layer.

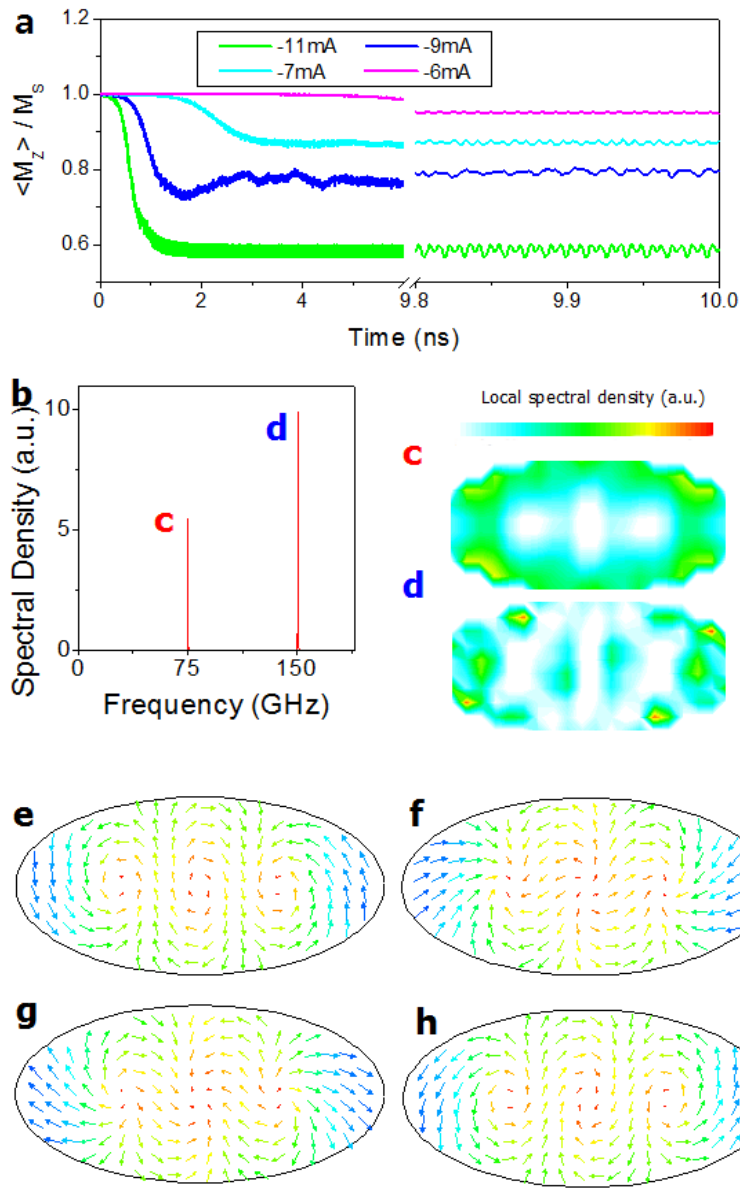

Fig. 6. Frequency and eigenmode analysis of current-induced excitation of single ferromagnetic layer: (a) Time evolution of the out-of-plane component of magnetization <$M_z$> at $\mu_0 H$ = 2.5 T and various negative currents. (b) Power spectrum at $\mu_0 H$ = 2.5 T and $I$ = –11 mA. (c) and (d) Eigenmode images for the two peak frequencies indicated in (b). (e)-(h) Magnetic domain patterns at various times after the onset of current: (e) 9.988 ns, (f) 9.992 ns, (g) 9.996 ns, and (h) 10.000 ns.

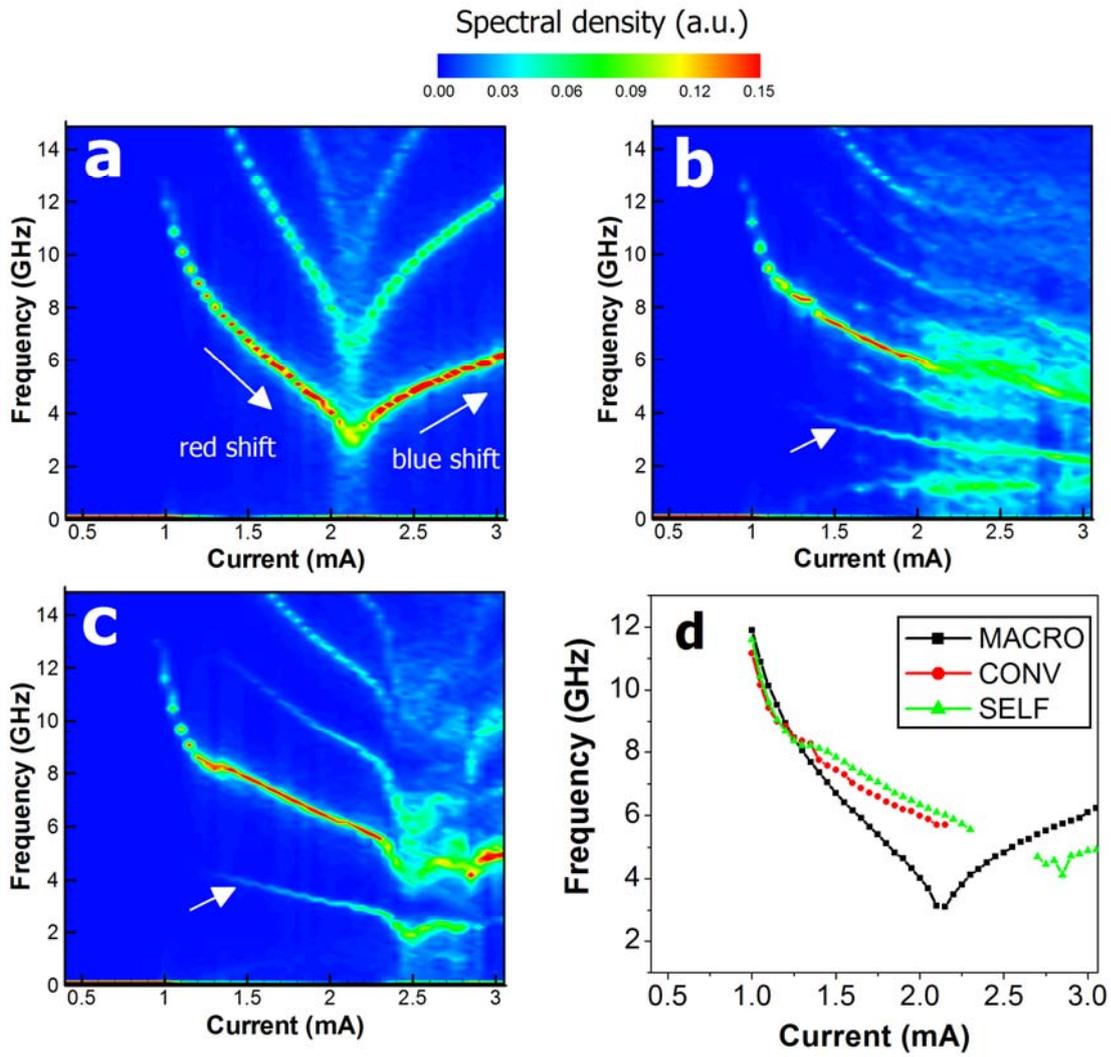

Fig. 7. Comparisons of spectral densities of a spin valve, obtained from three different models: (a) Macrospin model (MACRO), (b) Conventional micromagnetic model without considering non-local spin transfer torque (CONV), and (c) Self-consistent model (SELF). (d) Main peaks of the microwave oscillation obtained from the three models.

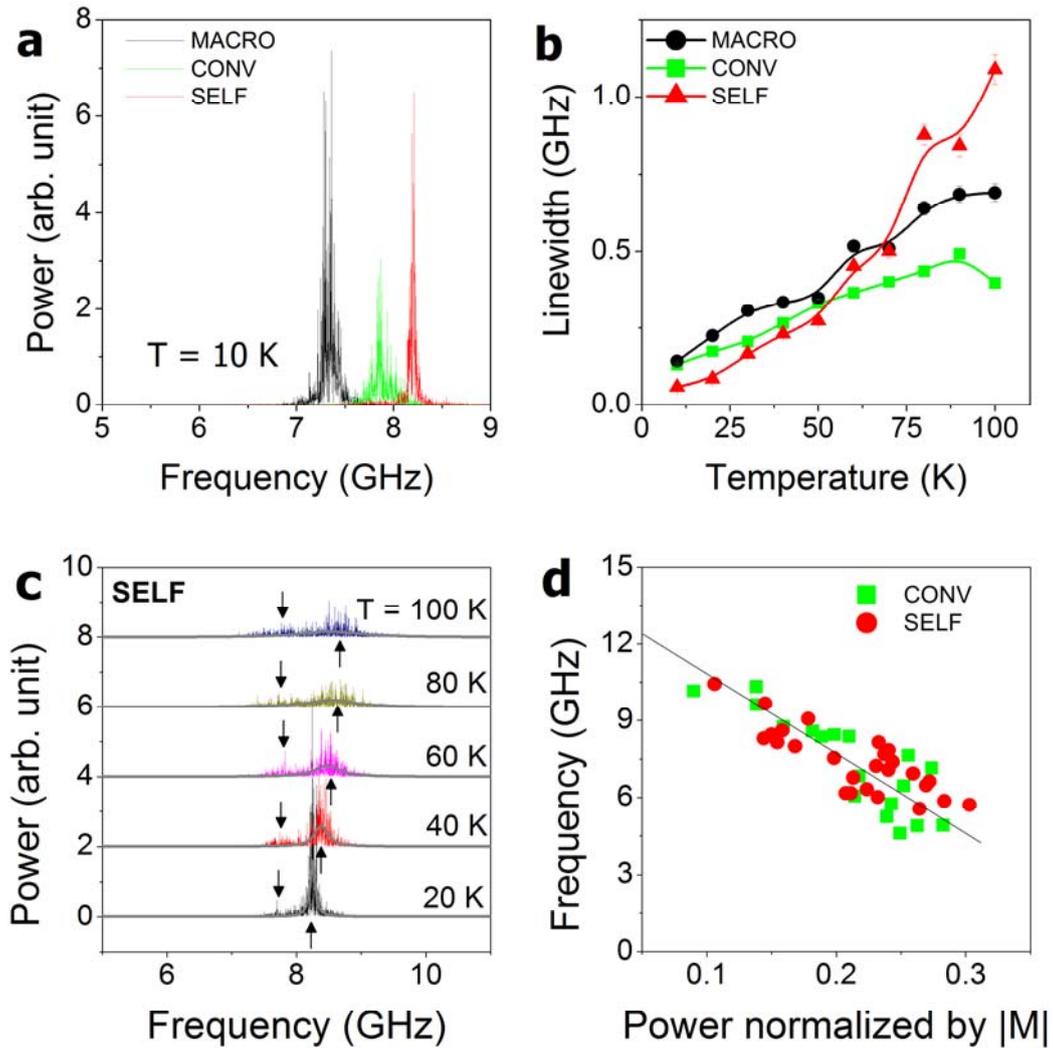

Fig. 8. Effects of non-local spin transfer torque on the linewidth. (a) Comparison of power spectra obtained in the three models at $T$ = 10 K. (b) Linewidth as a function of the temperature. (c) Power spectra obtained from non-local, self-consistent model as a function of the temperature. The spectra are vertically offset for clarity. Down-arrows indicate narrow secondary peaks whereas up-arrows indicate broad main peaks. Gray lines correspond to Lorentzian fits. (d) The frequency versus the power normalized by |**M**|.

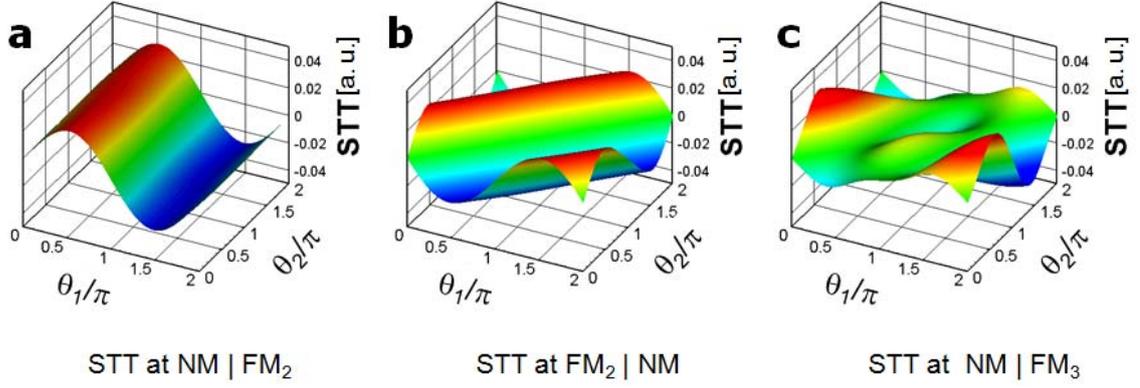

Fig. 9. Angular dependence of spin torque in a multilayer of NM(10) | FM$_1$(5) | NM(5) | FM$_2$(3) | NM (1) | FM$_3$(3) (all in nanometers). We assume the following spin transport parameters: For FM and NM, respectively, the parameters are bulk resistivity $\rho$ = 51 nΩ·m and 5 nΩ·m, bulk spin asymmetry $\beta_s$ = 0.51 and 0, spin diffusion length $l_{sf}$ = 60 nm and 1000 nm. For the interface parameters, FM | NM (or NM | FM), the parameters are interfacial resistance AR* = 0.52 fΩ·m$^2$, interfacial spin asymmetry $\gamma_s$ = 0, and spin mixing conductivity Re($G_{\uparrow\downarrow}$) = 5.42×10$^{14}$ Ω$^{-1}$m$^{-2}$. $\theta_1$ and $\theta_2$ represent the magnetization angles of FM$_2$ and FM$_3$ with respect to the magnetization angle of FM$_1$, respectively.

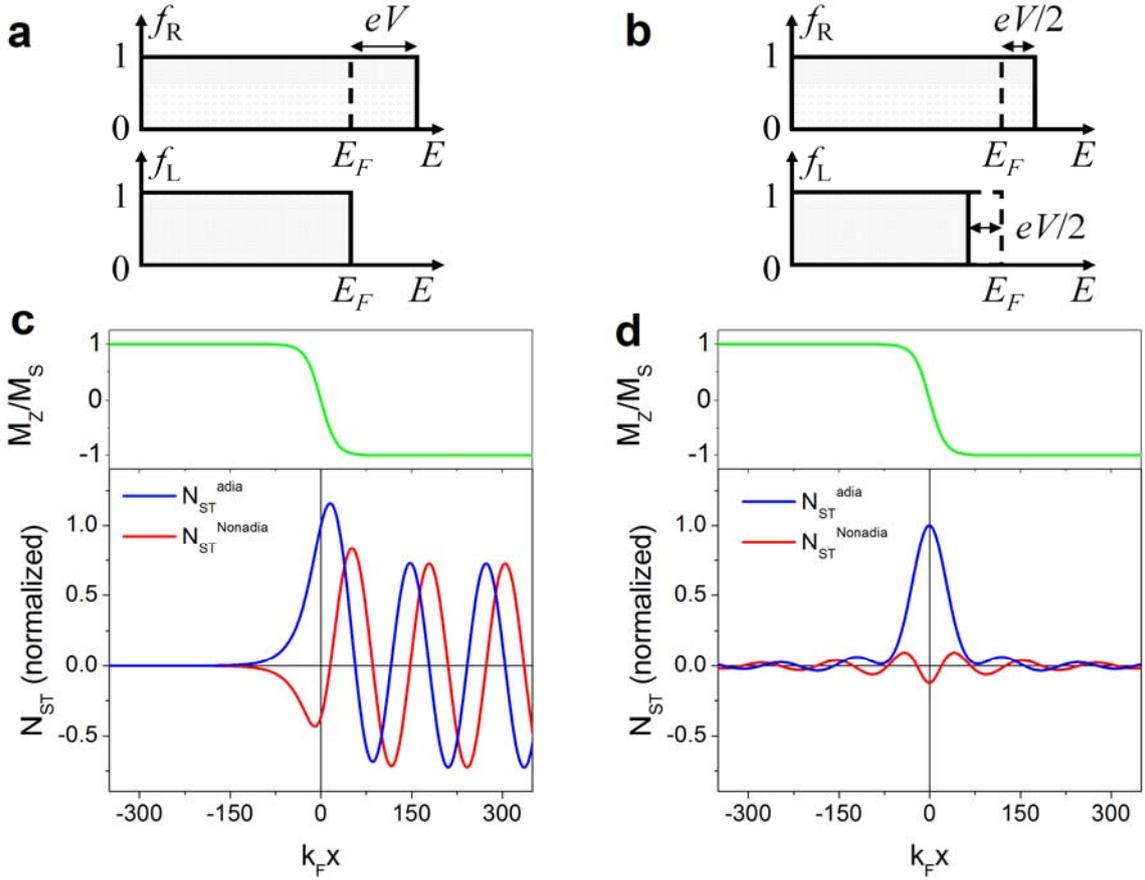

Fig. 10. Electron occupation probabilities $f_R$ and $f_L$ for the right- and left-propagating electrons as a function of the energy in (a) charge-neutrality-broken calculation and (b) charge-neutrality-preserved calculation. Spatial distribution of adiabatic ($\mathbf{N}_{ST}^{Adia}$) and nonadiabatic ($\mathbf{N}_{ST}^{Nonadia}$) spin transfer torques for a narrow domain wall centered at $x = 0$. (c) Charge-neutrality-broken calculation [116], and (d) charge-neutrality-preserved calculation (our work). Only the $k$-normal channel is considered ($\mathbf{k} = (+k_F, 0, 0)$) in (c), whereas the integration over the Fermi surface is performed in (d). Here, $K_u$ is assumed to be $4.5\times10^6$ J/m$^3$ and the upper panels of (c) and (d) show the domain wall profile.

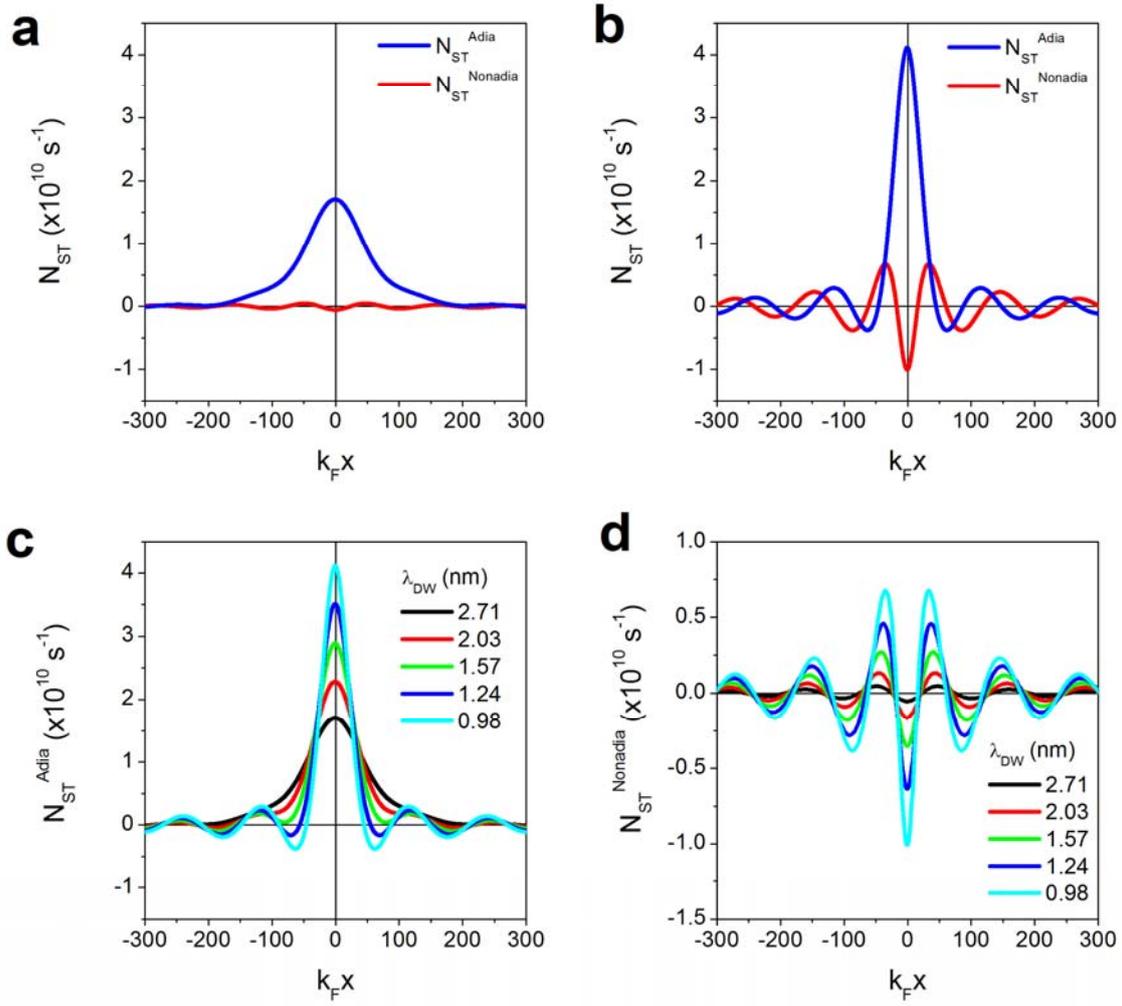

Fig. 11. Non-locality of spin transfer torque for narrow domain walls. Adiabatic ($\mathbf{N}_{ST}^{Adia}$) and nonadiabatic ($\mathbf{N}_{ST}^{Nonadia}$) vector components of spin transfer torque for (a) $K_u = 2 \times 10^6$ J/m$^3$ ($\lambda_{DW} \approx 2.71$ nm) and (b) $K_u = 10^7$ J/m$^3$ ($\lambda_{DW} \approx 0.98$ nm). (c) $\mathbf{N}_{ST}^{Adia}$ for various $\lambda_{DW}$ values. (d) $\mathbf{N}_{ST}^{Nonadia}$ for various $\lambda_{DW}$ values. Here, $J_e$ is $10^{12}$ A/m$^2$.

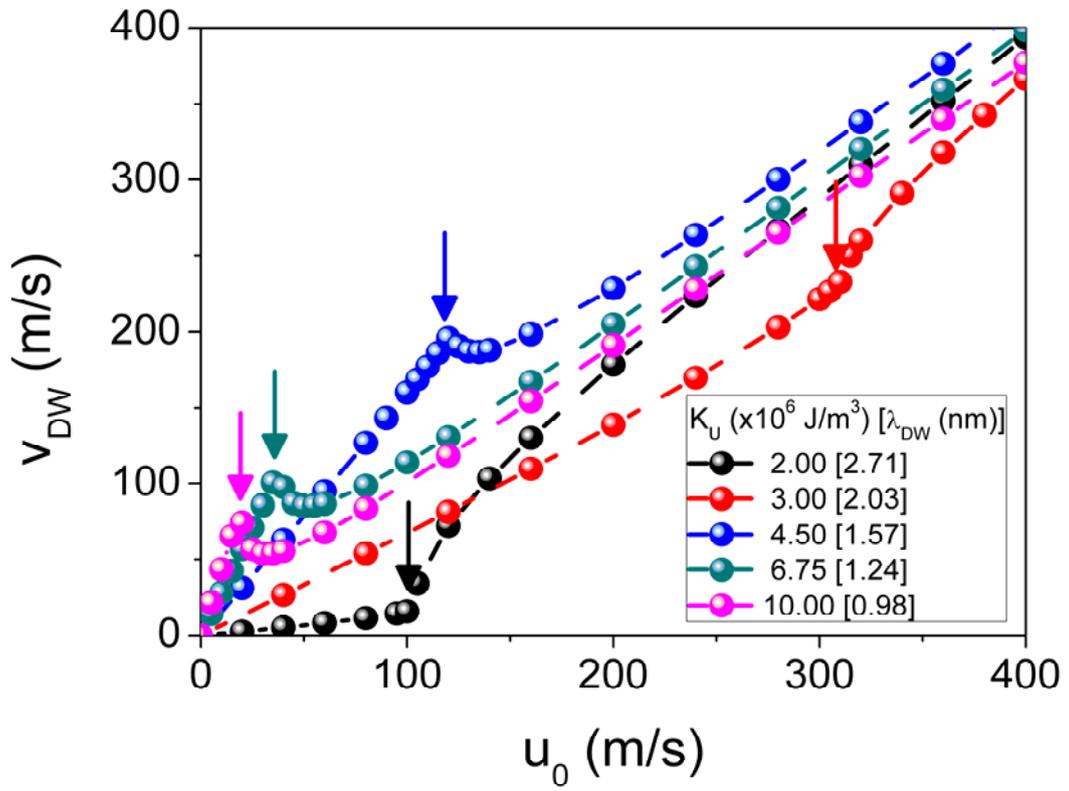

Fig. 12. Domain wall velocity ($v_{DW}$) versus spin current velocity ($u_0$) for various domain wall width ($\lambda_{DW}$).

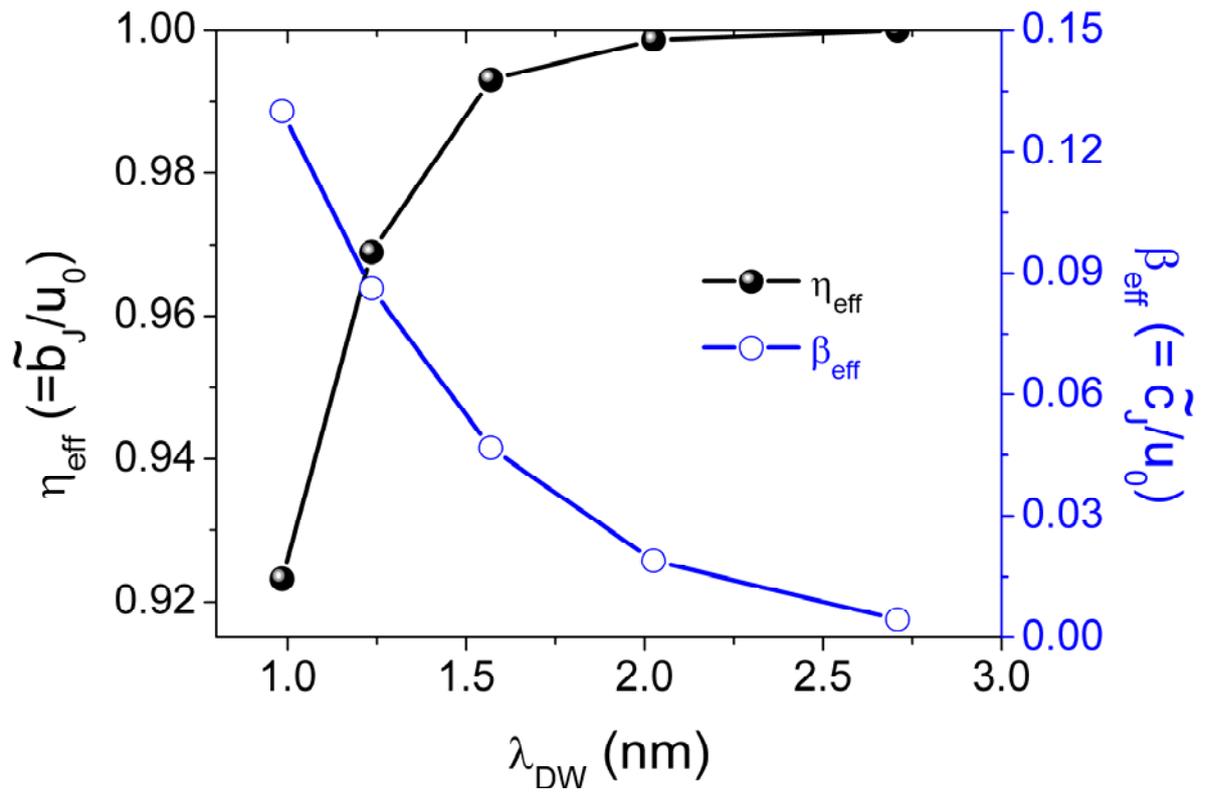

Fig. 13. Effective spin polarization ($\eta_{eff}$) and effective nonadiabaticity ($\beta_{eff}$) as a function of the domain wall width ($\lambda_{DW}$).

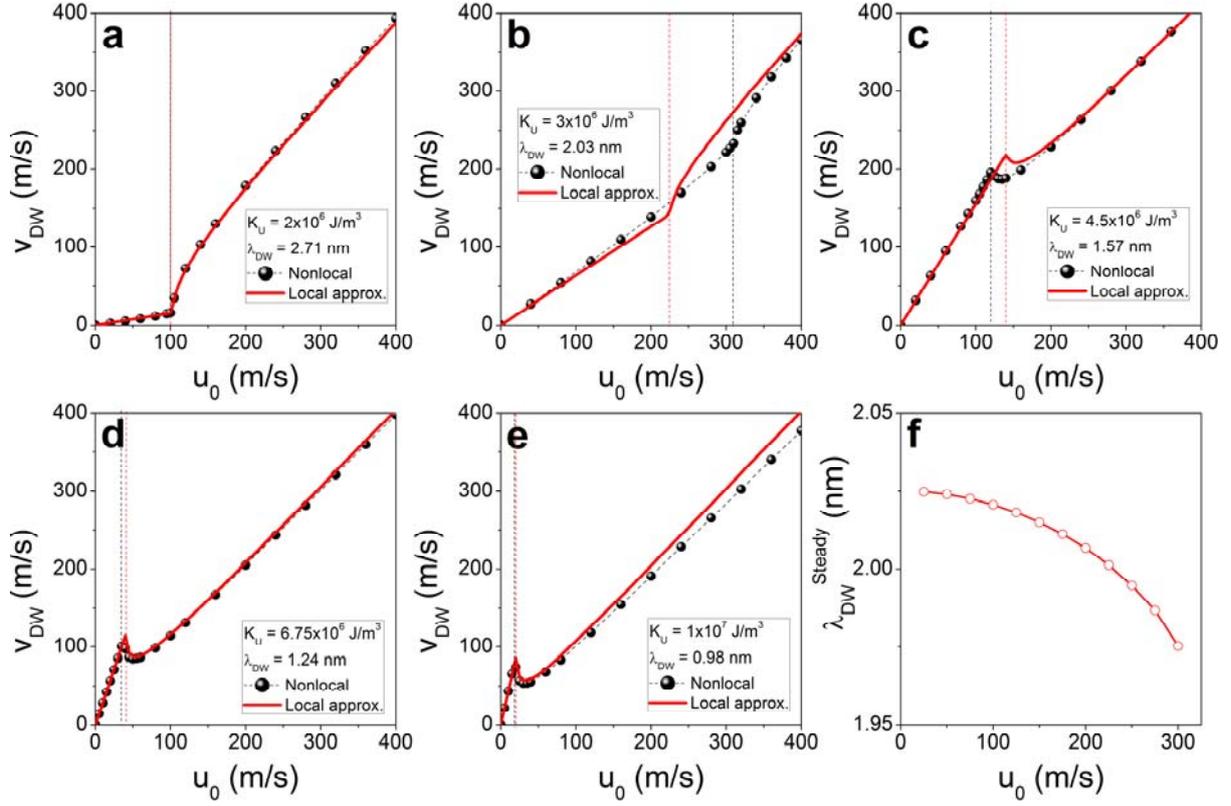

Fig. 14. Comparisons of domain wall velocities between calculations with local approximation (red solid lines) and self-consistent calculations (symbols). (a) $K_u = 2\times10^6$ J/m$^3$ ($\lambda_{DW}$ = 2.71 nm). (b) $K_u = 3\times10^6$ J/m$^3$ ($\lambda_{DW}$ = 2.03 nm). (c) $K_u = 4.5\times10^6$ J/m$^3$ ($\lambda_{DW}$ = 1.57 nm). (d) $K_u = 6.75\times10^6$ J/m$^3$ ($\lambda_{DW}$ = 1.24 nm). (e) $K_u = 1\times10^7$ J/m$^3$ ($\lambda_{DW}$ = 0.98 nm). (f) $\lambda_{DW}$ in the steady state ($\lambda_{DW}^{Steady} \equiv \lambda_{DW}$ at $t \to \infty$) versus $u_0$ for $K_u = 3\times10^6$ J/m$^3$. Vertical dotted lines correspond to the Walker breakdown thresholds.

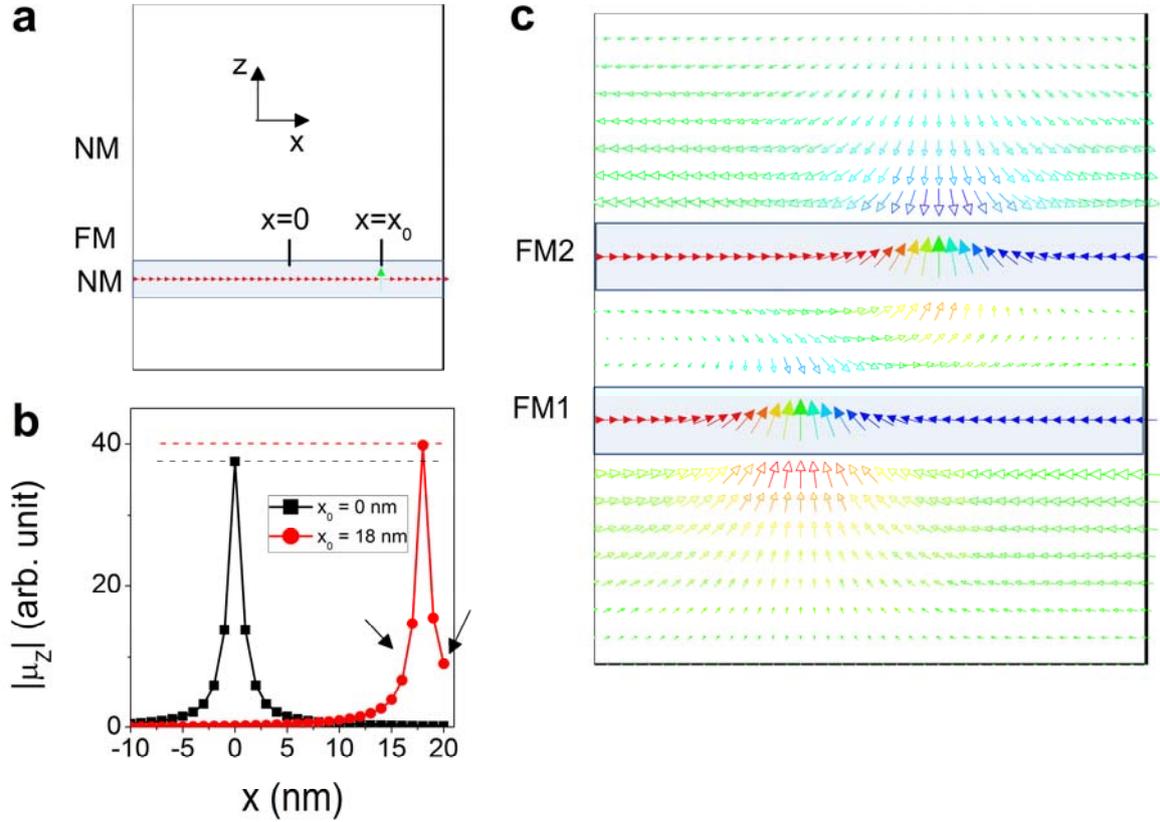

Fig. A1. Two tests of the diffusion kernel. (a) Schematic of a model system consisting of NM (10 nm) | FM (8 nm) | NM (32 nm). We assume that the magnetization is in the *x*-direction except for the cell located at $x = x_0$ is in *z* direction. (b) The *z*-component of spin chemical potential $\mu_z$, calculated by our approach, in the NM region at the bottom interface of FM|NM for two cases; $x_0 = 0$ (center of nanopillar) and $x_0 = 18$ nm (close to an edge). (c) Schematic of a model system consisting of NM (16 nm) | FM1 (6 nm) | NM (6 nm) | FM2 (6 nm) | NM (16 nm). In (c), arrows in the NM layers (hollow head) show the spin accumulation vectors and the arrows in the FM layers (filled head) show local magnetization vectors.